\documentclass[12pt]{article}

\usepackage{cite}
\usepackage{graphicx}
\usepackage{pict2e}
\usepackage[all]{xy}

\author{Yu.~M.~Zinoviev
       \thanks{E-mail address: Yurii.Zinoviev@ihep.ru} \\[0.5cm]
        {\it Institute for High Energy Physics} \\
        {\it of National Research Center "Kurchatov Institute"} \\
        {\it Protvino, Moscow Region, 142280, Russia}}

\title{Massive spin-2 supermultiplet and supergravity}

\date{}

\begin{document}

\maketitle

\begin{abstract}
In this work, we consider the interaction of massless $N=1$
supergravity with a massive $(2,3/2,3/2,1)$ supermultiplet, as a
possible candidate for a supersymmetric extension to bigravity. A
gauge invariant description for massive spin-2, spin-3/2 and spin-1
fields is used. As a result of this, ambiguities arise related to
possible field redefinitions, which are fixed by using a recently
proposed method based on unfolded equations, and by restricting
ourselves to interactions with the minimum number of derivatives (two
for bosonic and one for fermionic vertices). It appears that this
completely fixes the entire construction.
\end{abstract}

\thispagestyle{empty}
\newpage
\setcounter{page}{1}

\section{Introduction}

In this work, we investigate the interactions of a massive spin-2
supermultiplet with $N=1$ supergravity. On the one hand, this can be
considered as a modest first step towards the  general task of
constructing interactions between supergravities and massive 
higher-spin supermultiplets, which is important because any
superstring theory is simply a supergravity model with an infinite
number of such supermultiplets from a supersymmetric perspective.

On the other hand, a massive supermultiplet $(2,3/2,3/2,1)$ does
not contain fields with spins greater than 2, so it may be possible to
construct a closed theory with a finite number of components that
would be a minimal supersymmetric extension of Bigravity
\cite{HRS11,HR11}. Recall, that massive spin-2 supermultiplets exist
for $N=k$, $k=1,2,3,4$ supersymmetries. These supermultiplets were
explicitly constructed in \cite{Zin02}, where it was demonstrated that
in the massless limit they correctly reproduced the spectrum of $N=2k$
supergravities, and thus $N=4$ supergravity with a massive spin-2
supermultiplet can be considered a maximally supersymmetric extension
of Bigravity\footnote{See \cite{Tro26} for recent discussion of
supersymmetric extensions of usual gravity including the case of
spontaneously broken supergravities when massive spin-3/2
supermultiplets \cite{Zin07} appear.}.

As is well known, there are two types of $N=1$ massive supermultiplets
(with integer and half-integer superspins):
$$
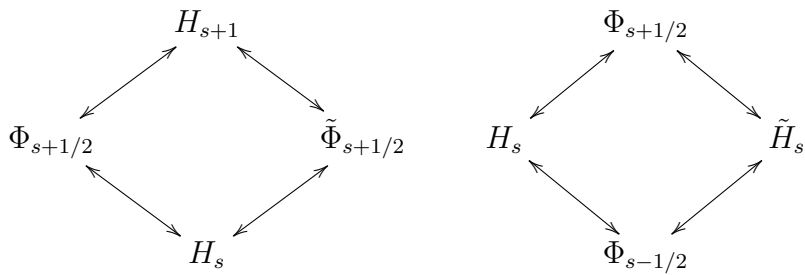
\xymatrix{
  & H_{s+1} \ar@{<->}[dl] &  &   & \Phi_{s+1/2} \ar@{<->}[dl] & \\
\Phi_{s+1/2} \ar@{<->}[dr] & & \tilde\Phi_{s+1/2} \ar@{<->}[ul] &
 H_s \ar@{<->}[dr] & & \tilde{H}_s \ar@{<->}[ul]  \\
  & H_s \ar@{<->}[ur] &    &  & \Phi_{s-1/2} \ar@{<->}[ur] }
$$
Besides the well-known fact that all members of a supermultiplet
must have equal masses, two more requirements are necessary for the
superalgebra to be closed: two bosons must have opposite parity,
while two fermions have opposite signs for their mass terms.
The explicit construction of a massive supermultiplets is a
 non-trivial task, even using powerful superfield techniques
(see e.g. \cite{Kou20,BFK26} and references therein), and till now,
the most effective way to create massive supermultiplets
\cite{Zin07a,BSZ17a,BKhSZ19,BKhSZ19a,BKhSZ19b,Zin23} is based
on the gauge invariant description of massive fields
\cite{Zin01,Met06,Zin08b,PV10,BSZ12a,BSZ14a,KhZ19}. In this approach a
massive field is described by a collection of massless fields
corresponding to each helicity, while supersymmetry connects
helicities that differ by half:
$$
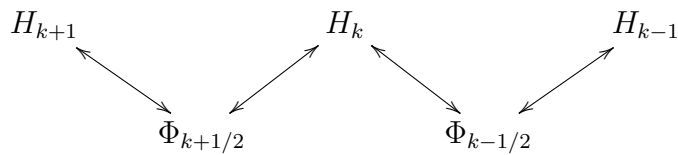
\xymatrix{
 H_{k+1} \ar@{<->}[dr] &   &  H_k \ar@{<->}[dr] &  & H_{k-1} \\
  &  \Phi_{k+1/2} \ar@{<->}[ur] &  &  \Phi_{k-1/2} \ar@{<->}[ur] &  }
$$

To construct interactions for massless fields and supermultiplets, one
can use the effective Fradkin-Vasiliev approach
\cite{FV87,FV87a,Vas11,BPS12,KhZ20a,KhZ20b}, based on the consistent
deformations of all gauge invariant objects (which we call
curvatures). However, when trying to apply this method to  massive
fields and supermultiplets using their gauge invariant description,
there are ambiguities due to field redefinitions caused by the
presence of Stueckelberg fields \cite{BDGT18,KhZ21,KhZ21a,Zin24a}. For
gravitational interactions of massive higher-spins fields, there are
well known substitution rules that allow one to directly obtain 
so-called minimal interactions. As for the non-minimal interactions
necessary for spins $s \ge 5/2$, they can be determined by restricting
the number of derivatives (\cite{Zin26} and references therein).
However, for supersymmetric models in the component formalism, there
are no simple rules for obtaining even minimal interactions.
Therefore, in this paper we use the following approach\footnote{For
the alternative approaches see e.g. \cite{MFG16,BCDL25}}. First, we
use a recently proposed method \cite{KhZ20,Zin24}, based on the
consistent deformation of unfolded equations, to  determine global
super transformations for massive fields. Then, by switching form
global to local supertransformations, we calculate minimal
interactions. Non-minimal interactions are fixed by requiring that
they must have the minimum possible number of derivatives, and this
completely determine the model.

The paper is organized as follows. Section 2 provides all necessary
kinematical information for the frame-like gauge invariant description
for the massive spin-2, spin-3/2 and spin-1 fields. Section 3 provides
a brief description of massless $N=1$ supergravity using our notation 
and conventions. Sections 4 and 5 discuss interactions with massless
gravitino and massless graviton respectively. Section 6 demonstrates
that our supertransformations correspond to a closed (on-shell)
superalgebra. Two appendixes describe the deformation procedure for
the unfolded equations.

\noindent
{\bf Notation and conventions} We work in  the frame-like multispinor
formalism where all objects are forms with some number of completely
symmetric dotted and un-dotted spinor indices $\alpha,\dot\alpha =
1,2$:
$$
\Phi^{(\alpha_1\dots\alpha_k)(\dot\alpha_1\dots\dot\alpha_l)}
= \Phi^{\alpha(k)\dot\alpha(l)}.
$$
A coordinate free description of the flat Minkowski space is based on
the background frame $e^{\alpha\dot\alpha}$ and background covariant
derivative $D$ such that
\begin{equation}
D \wedge e^{\alpha\dot\alpha} = 0, \qquad D \wedge D = 0.
\end{equation}
For the basis of forms we use two-forms $E^{\alpha(2)}$,
$E^{\dot\alpha(2)}$, three-form $E^{\alpha\dot\alpha}$ and four-form
$E$ defined as follows:
\begin{eqnarray}
e^{\alpha\dot\alpha} \wedge e^{\beta\dot\beta} &=& 
\epsilon^{\dot\alpha\dot\beta} E^{\alpha\beta} +
\epsilon^{\alpha\beta} E^{\dot\alpha\dot\beta}, \nonumber \\
E^{\alpha(2)} \wedge e^{\beta\dot\alpha} &=& \epsilon^{\alpha\beta}
E^{\alpha\dot\alpha}, \\
E^{\alpha\dot\alpha} \wedge e^{\beta\dot\beta} &=&
\epsilon^{\alpha\beta} \epsilon^{\dot\alpha\dot\beta} E. \nonumber
\end{eqnarray}
In what follows $\wedge$ signs are omitted.

\section{Kinematics}

In this section we provide all necessary kinematical information for
all members of massive spin 2 supermultiplet.

\subsection{Massive spin 2}

A frame-like, gauge invariant description of massive spin-2 requires
three physical fields: the one-forms $H^{\alpha\dot\alpha}$, $A$ and
the zero-form $\varphi$, and three auxiliary fields: the one-form
$\Omega^{\alpha(2)} + h.c.$ and the two zero-forms
$B^{\alpha(2)} + h.c.$, $\pi^{\alpha\dot\alpha}$, which are needed to
construct the Lagrangian of first order in derivatives. The free
Lagrangian looks like:
\begin{eqnarray}
\frac{1}{i} {\cal L}_0 &=& 2 \Omega^{\alpha\beta} E_\beta{}^\gamma
\Omega_{\alpha\gamma} + 2 \Omega^{\alpha\beta} e_\beta{}^{\dot\alpha}
D H_{\alpha\dot\alpha} + 8 E B_{\alpha(2)} B^{\alpha(2)} + 4
E_{\alpha(2)} B^{\alpha(2)} D A + h.c. \nonumber \\
 && - 12 E \phi_{\alpha\dot\alpha} \pi^{\alpha\dot\alpha} -
24 E_{\alpha\dot\alpha} \pi^{\alpha\dot\alpha} D \varphi \nonumber \\
 && - 4M [ \Omega^{\alpha(2)} E_{\alpha(2)} A - 2 B^{\alpha\beta}
E_\beta{}^{\dot\alpha} H_{\alpha\dot\alpha} + h.c.]
- 24M E_{\alpha\dot\alpha} \pi^{\alpha\dot\alpha} A \nonumber \\
 && + 2M^2 [ H^{\alpha\dot\alpha} E_\alpha{}^\beta
H_{\beta\dot\alpha} + h.c. ] + 24M^2 E_{\alpha\dot\alpha}
H^{\alpha\dot\alpha} \varphi + 48M^2 E \varphi^2.
\end{eqnarray}
The structure of the Lagrangian is typical of the gauge invariant
description of massive fields: it is a sum of the kinetic and
mass-like terms for all helicities, plus cross-terms that glue them
together. This Lagrangian is invariant under the following gauge
transformations:
\begin{eqnarray}
\delta \Omega^{\alpha(2)} &=& D \eta^{\alpha(2)} + \frac{M^2}{2}
e^\alpha{}_{\dot\alpha} \xi^{\alpha\dot\alpha}, \nonumber \\
\delta H^{\alpha\dot\alpha} &=& D \xi^{\alpha\dot\alpha} +
e_\beta{}^{\dot\alpha} \eta^{\alpha\beta} + e^\alpha{}_{\dot\beta}
\eta^{\dot\alpha\dot\beta} + M e^{\alpha\dot\alpha} \xi, \\
\delta B^{\alpha(2)} &=& - M \eta^{\alpha(2)}, \qquad
\delta A = D \xi + M e_{\alpha\dot\alpha} \xi^{\alpha\dot\alpha},
\nonumber \\
\delta \pi^{\alpha\dot\alpha} &=& - M^2 \xi^{\alpha\dot\alpha}, \qquad
\delta \varphi = - M \xi. \nonumber 
\end{eqnarray}
One of the nice features of this formalism is that each field has its
own gauge invariant object, which we call curvature (two-forms for
the gauge fields and one-forms for the Stueckelberg  fields):
\begin{eqnarray}
{\cal R}^{\alpha(2)} &=& D \Omega^{\alpha(2)} + \frac{M^2}{2}
e^\alpha{}_{\dot\beta} H^{\alpha\dot\beta} + M E^\alpha{}_\beta
B^{\alpha\beta} + 2M^2 E^{\alpha(2)} \varphi, \nonumber \\
{\cal T}^{\alpha\dot\alpha} &=& DH^{\alpha\dot\alpha} +
e_\beta{}^{\dot\alpha} \Omega^{\alpha\beta} + e^\alpha{}_{\dot\beta}
\Omega^{\dot\alpha\dot\beta} + M e^{\alpha\dot\alpha} A, \nonumber \\
{\cal B}^{\alpha(2)} &=& DB^{\alpha(2)} + M \Omega^{\alpha(2)} +
\frac{M}{2} e^\alpha{}_{\dot\beta}\pi^{\alpha\dot\beta}, \label{cur_2}
\\
{\cal A} &=& DA + 2(E_{\alpha(2)} B^{\alpha(2)} +
E_{\dot\alpha(2)}B^{\dot\alpha(2)}) + M e_{\alpha\dot\alpha}
H^{\alpha\dot\alpha}, \nonumber \\
\Pi^{\alpha\dot\alpha} &=& D\pi^{\alpha\dot\alpha}
+ M^2 H^{\alpha\dot\alpha} + M (e_\beta{}^{\dot\alpha} B^{\alpha\beta}
+ e^\alpha{}_{\dot\beta} B^{\dot\alpha\dot\beta}) +
M^2 e^{\alpha\dot\alpha} \varphi, \nonumber \\
{\cal C} &=& D\varphi + e_{\alpha\dot\alpha} \pi^{\alpha\dot\alpha}
+ M A. \nonumber 
\end{eqnarray}
In  what follows we use a term "partially on-shell" (an analogue of
the well known torsion zero condition in gravity) meaning that
\begin{equation}
{\cal T}^{\alpha\dot\alpha} \approx 0, \qquad
{\cal A} \approx 0, \qquad
{\cal C} \approx 0
\end{equation}
In this case the remaining curvatures satisfy a number of differential
constraints:
\begin{eqnarray}
D {\cal R}^{\alpha(2)} &\approx& M E^\alpha{}_\beta 
{\cal B}^{\alpha\beta}, \nonumber \\
D {\cal B}^{\alpha(2)} &\approx& M {\cal R}^{\alpha(2)} - \frac{M}{2}
e^\alpha{}_{\dot\alpha} \Pi^{\alpha\dot\alpha}, \\
D \Pi^{\alpha\dot\alpha} &\approx& - M (e_\beta{}^{\dot\alpha}
{\cal B}^{\alpha\beta} + e^\alpha{}_{\dot\beta} 
{\cal B}^{\dot\alpha\dot\beta}), \nonumber
\end{eqnarray}
and algebraic constraints
\begin{eqnarray}
0 &\approx& e_\beta{}^{\dot\alpha} {\cal R}^{\alpha\beta}
+ e^\alpha{}_{\dot\beta} {\cal R}^{\dot\alpha\dot\beta}, \nonumber \\
0 &\approx& E_{\alpha(2)} {\cal B}^{\alpha(2)} + E_{\dot\alpha(2)}
{\cal B}^{\dot\alpha(2)}, \\
0 &\approx& e_{\alpha\dot\alpha} \Pi^{\alpha\dot\alpha} \nonumber
\end{eqnarray}
Variation of the free Lagrangian under arbitrary variations of the
physical fields can be calculated as follows:
\begin{equation}
\frac{1}{i} \delta {\cal L}_0 = 2 [ {\cal R}_{\alpha\beta}
e^\beta{}_{\dot\alpha} - {\cal R}_{\dot\alpha\dot\beta}
e_\alpha{}^{\dot\beta}] \delta H^{\alpha\dot\alpha} - 4 
[E_{\alpha(2)} {\cal B}^{\alpha(2)} - E_{\dot\alpha(2)} 
{\cal B}^{\dot\alpha(2)}] \delta A + 24 E_{\alpha\dot\alpha}
\Pi^{\alpha\dot\alpha} \delta \varphi. 
\end{equation}

Now let us turn to the unfolded equations. For the physical fields
we have:
\begin{eqnarray}
0 &=& DH^{\alpha\dot\alpha} + e_\beta{}^{\dot\alpha}
\Omega^{\alpha\beta} + e^\alpha{}_{\dot\beta}
\Omega^{\dot\alpha\dot\beta} + M e^{\alpha\dot\alpha} A, \nonumber \\
0 &=& DA + 2(E_{\alpha(2)} B^{\alpha(2)} + E_{\dot\alpha(2)}
B^{\dot\alpha(2)}) + M e_{\alpha\dot\alpha} H^{\alpha\dot\alpha}, \\
0 &=& D\varphi + e_{\alpha\dot\alpha} \pi^{\alpha\dot\alpha} + M A,
\nonumber 
\end{eqnarray}
and for the auxiliary fields:
\begin{eqnarray}
0 &=& D \Omega^{\alpha(2)} + \frac{M^2}{2}
e^\alpha{}_{\dot\beta} H^{\alpha\dot\beta} + M E^\alpha{}_\beta
B^{\alpha\beta} + 2M^2 E^{\alpha(2)} \varphi - E_{\beta(2)}
 W^{\alpha(2)\beta(2)}, \nonumber \\
0 &=& DB^{\alpha(2)} + M \Omega^{\alpha(2)} + \frac{M}{2} 
e^\alpha{}_{\dot\beta}\pi^{\alpha\dot\beta} - e_{\beta\dot\alpha}
 B^{\alpha(2)\beta\dot\alpha}, \\
0 &=& D\pi^{\alpha\dot\alpha} + M^2 H^{\alpha\dot\alpha} + M
(e_\beta{}^{\dot\alpha} B^{\alpha\beta} + e^\alpha{}_{\dot\beta}
B^{\dot\alpha\dot\beta}) + M^2 e^{\alpha\dot\alpha} \varphi 
- e_{\beta\dot\beta} \pi^{\alpha\beta\dot\alpha\dot\beta}. \nonumber
\end{eqnarray}
Here $W^{\alpha(4)}$, $B^{\alpha(3)\dot\alpha}$, 
$\pi^{\alpha(2)\dot\alpha(2)}$ are just the first representatives of
infinite chains of gauge invariant zero-forms satisfying the
equations $(0 \le k < \infty)$:
\begin{eqnarray}
0 &=& D W^{\alpha(4+k)\dot\alpha(k)} - e_{\beta\dot\beta}
W^{\alpha(4+k)\beta\dot\alpha(k)\dot\beta} + a_{1,k}
e^\alpha{}_{\dot\beta} B^{\alpha(3+k)\dot\alpha(k)\dot\beta}
+ b_{1,k} e^{\alpha\dot\alpha} W^{\alpha(3+k)\dot\alpha(k-1)},
\nonumber \\
0 &=&  B^{\alpha(3+k)\dot\alpha(k+1)} - e_{\beta\dot\beta}
B^{\alpha(3+k)\beta\dot\alpha(k+1)\dot\beta} + a_{2,k}
e_\beta{}^{\dot\alpha} W^{\alpha(3+k)\beta\dot\alpha(k)} \nonumber \\
 && + a_{3,k} e^\alpha{}_{\dot\beta} 
\pi^{\alpha(2+k)\dot\alpha(k+1)\dot\beta} + b_{2,k}
e^{\alpha\dot\alpha} B^{\alpha(2+k)\dot\alpha(k)}, \\
0 &=& D \pi^{\alpha(2+k)\dot\alpha(2+k)} - e_{\beta\dot\beta}
\pi^{\alpha(2+k)\beta\dot\alpha(2+k)\dot\beta} + a_{4,k}
e_\beta{}^{\dot\alpha} B^{\alpha(2+k)\beta\dot\alpha(k+1)} \nonumber 
\\
 && + a_{4,k} e^\alpha{}_{\dot\beta} 
B^{\alpha(1+k)\dot\alpha(k+2)\dot\beta} + b_{3,k} e^{\alpha\dot\alpha}
\pi^{\alpha(1k)\dot\alpha(k+1)}. \nonumber 
\end{eqnarray}
$$
a_{1,k} = \frac{4M}{(k+4)(k+5)}, \qquad
a_{2,k} = \frac{M}{(k+1)(k+2)},
$$
$$
a_{3,k} = \frac{3M}{(k+3)(k+4)}, \qquad
a_{4,k} = \frac{2M}{(k+2)(k+3)},
$$
$$
b_{1,k} = \frac{M^2}{(k+1)(k+4)}, \qquad
b_{3,k} = \frac{k(k+5)M^2}{(k+2)^2(k+3)^2},
$$
$$
b_{2,k} = \frac{k(k+5)M^2}{(k+1)(k+2)(k+3)(k+4)},
$$

\subsection{Massive spin 3/2}

In this case we need just a couple of physical fields: one-form
$\Phi^\alpha + h.c.$ and zero-from $\phi^\alpha + h.c.$. The free
Lagrangian has the form:
\begin{eqnarray}
{\cal L}_0 &=& - \Phi_\alpha e^\alpha{}_{\dot\alpha} D
\Phi^{\dot\alpha} - 6 \phi_\alpha E^\alpha{}_{\dot\alpha} D
\phi^{\dot\alpha} \nonumber \\
 && - \epsilon M \Phi_\alpha E^\alpha{}_\beta \Phi^\beta - 6M
\Phi_\alpha E^\alpha{}_{\dot\alpha} \phi^{\dot\alpha} 
+ 6\epsilon M E \phi_\alpha \phi^\alpha + h.c.
\end{eqnarray}
where $\epsilon = \pm 1$. This Lagrangian is invariant under the
following gauge transformations:
\begin{equation}
\delta \Phi^\alpha = D \rho^\alpha + \epsilon M 
e^\alpha{}_{\dot\alpha} \rho^{\dot\alpha}, \qquad 
\delta \phi^\alpha = - M \rho^\alpha
\end{equation}
Both fields has its own gauge invariant curvatures:
\begin{eqnarray}
{\cal F}^\alpha &=& D \Phi^\alpha + \epsilon M e^\alpha{}_{\dot\alpha}
\Phi^{\dot\alpha} + 2M E^\alpha{}_\beta \phi^\beta, \nonumber \\
{\cal C}^\alpha &=& D \phi^\alpha + M \Phi^\alpha + \epsilon M
e^\alpha{}_{\dot\alpha} \phi^{\dot\alpha}. 
\end{eqnarray}
Using them one can calculate a variation of the free Lagrangian under
the arbitrary variations of the physical fields as follows:
\begin{equation}
\delta {\cal L}_0 = - {\cal F}_{\dot\alpha} e_\alpha{}^{\dot\alpha}
\delta \Phi^\alpha + {\cal F}_\alpha e^\alpha{}_{\dot\alpha} \delta
 \Phi^{\dot\alpha} - 6 {\cal C}_{\dot\alpha} E_\alpha{}^{\dot\alpha}
\delta \phi^\alpha - 6 {\cal C}_\alpha E^\alpha{}_{\dot\alpha}
\delta \phi^{\dot\alpha}.
\end{equation}

The unfolded equations for the physical fields have the form:
\begin{eqnarray}
0 &=& D \Phi^\alpha + \epsilon M e^\alpha{}_{\dot\alpha}
\Phi^{\dot\alpha} + 2M E^\alpha{}_\beta \phi^\beta - E_{\beta(2)}
Y^{\alpha\beta(2)}, \nonumber \\
0 &=& D \phi^\alpha + M \Phi^\alpha + \epsilon M 
e^\alpha{}_{\dot\alpha} \phi^{\dot\alpha} - e_{\beta\dot\alpha}
\phi^{\alpha\beta\dot\alpha}, 
\end{eqnarray}
where $Y^{\alpha(3)}$ and $\phi^{\alpha(2)\dot\alpha}$ are just the
first representatives of infinite chains of gauge invariant zero-forms
satisfying the equations $(0 \le k < \infty)$:
\begin{eqnarray}
0 &=& D Y^{\alpha(3+k)\dot\alpha(k)} - e_{\beta\dot\beta}
Y^{\alpha(3+k)\beta\dot\alpha(k)\dot\beta} + c_{1,k}
e^\alpha{}_{\dot\beta} \phi^{\alpha(2k)\dot\alpha(k)\dot\beta}
+ d_{1,k} e^{\alpha\dot\alpha} Y^{\alpha(2+k)\dot\alpha(k-1)},
\nonumber \\
0 &=& D \phi^{\alpha(2+k)\dot\alpha(k+1)} - e_{\beta\dot\beta}
\phi^{\alpha(2+k)\beta\dot\alpha(k+1)\dot\beta} + c_{2,k}
e_\beta{}^{\dot\alpha} Y^{\alpha(2+k)\beta\dot\alpha(k)} \\
 && + c_{3,k} e^\alpha{}_{\dot\beta} 
\phi^{\alpha(1+k)\dot\alpha(k+1)\dot\beta} + d_{2,k}
e^{\alpha\dot\alpha} \phi^{\alpha(1+k)\dot\alpha(k)}, \nonumber
\end{eqnarray}
$$
c_{1,k} = \frac{3M}{(k+3)(k+4)}, \qquad
c_{2,k} = \frac{M}{(k+1)(k+2)}, \qquad
c_{3,k} = \frac{2\epsilon M}{(k+2)(k+3)}
$$
$$
d_{1,k} = \frac{M^2}{(k+1)(k+3)}, \qquad
d_{2,k} = \frac{k(k+4)M^2}{(k+1)(k+2)^2(k+3)}
$$

\subsection{Massive spin 1}

A frame-like gauge invariant  description of massive spin-1 requires a
couple of physical fields: one-form $\tilde{A}$ and zero-form 
$\tilde\varphi$, and two auxiliary fields: zero-forms 
$\tilde{B}^{\alpha(2)} + h.c.$ and $\tilde\pi^{\alpha\dot\alpha}$.
The free Lagrangian looks like:
\begin{eqnarray}
\frac{1}{i} {\cal L}_0 &=& 2 E [\tilde{B}_{\alpha(2)}
\tilde{B}^{\alpha(2)} + \tilde{B}_{\dot\alpha(2)}
\tilde{B}^{\dot\alpha(2)}] + [E_{\alpha(2)} \tilde{B}^{\alpha(2)} -
E_{\dot\alpha(2)} \tilde{B}^{\dot\alpha(2)}] D\tilde{A} \nonumber \\
 && - 4E \tilde\pi_{\alpha\dot\alpha} \tilde\pi^{\alpha\dot\alpha} - 8
E_{\alpha\dot\alpha} \tilde\pi^{\alpha\dot\alpha} D \tilde\varphi - 4M
E_{\alpha\dot\alpha} \tilde\pi^{\alpha\dot\alpha} \tilde{A} 
\end{eqnarray}
This Lagrangian is invariant under the following gauge transformations
\begin{equation}
\delta \tilde{A} = D \tilde\xi, \qquad \delta \tilde\varphi = -
\frac{M}{2} \tilde\xi
\end{equation}
Gauge invariant curvatures for all fields have the form:
\begin{eqnarray}
\tilde{\cal B}^{\alpha(2)} &=& D \tilde{B}^{\alpha(2)} + \frac{M}{3} 
e^\alpha{}_{\dot\alpha} \tilde\pi^{\alpha\dot\alpha}, \nonumber \\
\tilde\Pi^{\alpha\dot\alpha} &=& D \tilde\pi^{\alpha\dot\alpha} +
\frac{M}{2} (e_\beta{}^{\dot\alpha} \tilde{B}^{\alpha\beta} + 
e^\alpha{}_{\dot\beta} \tilde{B}^{\dot\alpha\dot\beta}), \nonumber \\ 
\tilde{\cal A} &=& D \tilde{A} + 2(E_{\alpha(2)} \tilde{B}^{\alpha(2)}
+ E_{\dot\alpha(2)} \tilde{B}^{\dot\alpha(2)}), \label{cur_1} \\
\tilde{\cal C} &=& D \tilde\varphi + e_{\alpha\dot\alpha}
\tilde\pi^{\alpha\dot\alpha} + \frac{M}{2} \tilde{A}. \nonumber 
\end{eqnarray}
In this case partially on-shell means that
\begin{equation}
\tilde{\cal A} \approx 0, \qquad \tilde{\cal C} \approx 0,
\end{equation}
while the remaining curvatures satisfy the algebraic constraints
\begin{equation}
E_{\alpha(2)} \tilde{\cal B}^{\alpha(2)} + E_{\dot\alpha(2)}
\tilde{\cal B}^{\dot\alpha(2)} \approx 0, \qquad
e_{\alpha\dot\alpha} \tilde\Pi^{\alpha\dot\alpha} \approx 0.
\end{equation}
Variation of the free Lagrangian under the arbitrary variations of the
physical fields can be calculated as follows:
\begin{equation}
\frac{1}{i} \delta {\cal L}_0 = - (E_{\alpha(2)} 
\tilde{\cal B}^{\alpha(2)} - E_{\dot\alpha(2)} 
\tilde{\cal B}^{\dot\alpha(2)}) \delta \tilde{A} + 8
E_{\alpha\dot\alpha} \tilde\Pi^{\alpha\dot\alpha} \delta
\tilde\varphi.
\end{equation}

The unfolded equations for the physical fields have the form
\begin{eqnarray}
0 &=& D \tilde{A} + 2(E_{\alpha(2)} \tilde{B}^{\alpha(2)} +
E_{\dot\alpha(2)} \tilde{B}^{\dot\alpha(2)}), \nonumber \\
0 &=& D \tilde\varphi + e_{\alpha\dot\alpha}
\tilde\pi^{\alpha\dot\alpha} + \frac{M}{2} \tilde{A}, 
\end{eqnarray}
and for the auxiliary fields we have:
\begin{eqnarray}
0 &=& D \tilde{B}^{\alpha(2)} + \frac{M}{3} e^\alpha{}_{\dot\alpha}
\tilde\pi^{\alpha\dot\alpha} - e_{\beta\dot\beta} 
\tilde{B}^{\alpha(2)\beta\dot\beta}, \nonumber \\
0 &=& D \tilde\pi^{\alpha\dot\alpha} + \frac{M}{2} 
(e_\beta{}^{\dot\alpha} \tilde{B}^{\alpha\beta} +  
e^\alpha{}_{\dot\beta} \tilde{B}^{\dot\alpha\dot\beta})  -
e_{\beta\dot\beta} \tilde\pi^{\alpha\beta\dot\alpha\dot\beta}, 
\end{eqnarray}
where $\tilde{B}^{\alpha(3)\dot\alpha}$ and 
$\tilde\pi^{\alpha(2)\dot\alpha(2}$ are just the first representatives
of infinite chains of gauge invariant zero-forms satisfying the
equations $(0 \le k < \infty)$:
\begin{eqnarray}
0 &=& D \tilde{B}^{\alpha(2+k)\dot\alpha(k)} - e_{\beta\dot\beta}
\tilde{B}^{\alpha(2+k)\beta\dot\alpha(k)\dot\beta} + a_{1,k}
e^\alpha{}_{\dot\beta} \tilde\pi^{\alpha(1+k)\dot\alpha(k)\dot\beta} +
b_{1,k} e^{\alpha\dot\alpha} \tilde{B}^{\alpha(1+k)\dot\alpha(k-1)},
\nonumber \\
0 &=& D \tilde\pi^{\alpha(1+k)\dot\alpha(1+k)} - e_{\beta\dot\beta}
\tilde\pi^{\alpha(1+k)\beta\dot\alpha(1+k)\dot\beta} + a_{2,k}
e_\beta{}^{\dot\alpha} \tilde{B}^{\alpha(1+k)\beta\dot\alpha(k)} \\
 && + a_{2,k} e^\alpha{}_{\dot\beta} 
\tilde{B}^{\alpha(1+k)\dot\alpha(1+k)\dot\beta} + b_{2,k}
e^{\alpha\dot\alpha} \tilde\pi^{\alpha(k)\dot\alpha(k)}, \nonumber 
\end{eqnarray}
$$
a_{1,k} = \frac{2M}{(k+2)(k+3)}, \qquad
a_{2,k} = \frac{M}{(k+1)(k+2)}
$$
$$
b_{1,k} = \frac{M^2}{(k+1)(k+2)}, \qquad
b_{2,k} = \frac{k(k+3)M^2}{(k+1)^2(k+2)^2}
$$

\section{$N=1$ supergravity}

For the massless spin-2 we use $h^{\alpha\dot\alpha}$, 
$\omega^{\alpha(2)} + h.c.$ with the free Lagrangian
\begin{equation}
\frac{1}{i}{\cal L}_0 = - 2 \omega^\alpha{}_\beta E^{\beta\gamma}
\omega_{\alpha\gamma} - 2 D \omega^{\alpha\beta} 
e_\beta{}^{\dot\alpha} h_{\alpha\dot\alpha} + h.c.
\end{equation}
gauge transformations:
\begin{equation}
\delta \omega^{\alpha(2)} = D \hat\eta^{\alpha(2)}, \qquad
\delta h^{\alpha\dot\alpha} = D \hat\xi^{\alpha\dot\alpha} +
e_\beta{}^{\dot\alpha} \hat\eta^{\alpha\beta} + e^\alpha{}_{\dot\beta}
\hat\eta^{\dot\alpha\dot\beta}
\end{equation}
and gauge invariant curvatures
\begin{equation}
R^{\alpha(2)} = D \omega^{\alpha(2)}, \qquad
T^{\alpha\dot\alpha} = D h^{\alpha\dot\alpha} + e_\beta{}^{\dot\alpha}
\omega^{\alpha\beta} + e^\alpha{}_{\dot\beta}
\omega^{\dot\alpha\dot\beta} 
\end{equation}
For massless gravitino we use $\Psi^\alpha + h.c.$ with the free
Lagrangian
\begin{equation}
{\cal L}_0 = - \frac{1}{2} \Psi_\alpha e^\alpha{}_{\dot\alpha}
D \Psi^{\dot\alpha} + \frac{1}{2} \Psi_{\dot\alpha} 
e_\alpha{}^{\dot\alpha} D \Psi^\alpha
\end{equation}
gauge transformations and gauge invariant curvature
\begin{equation}
\delta \Psi^\alpha = D \zeta^\alpha, \qquad
{\cal F}^\alpha = D \Psi^\alpha
\end{equation}
Cubic vertex (we set coupling constant to be 1):
\begin{eqnarray}
{\cal L}_1 &=& - 2 \omega^\alpha{}_\gamma \omega^{\beta\gamma} 
e_\beta{}^{\dot\alpha} h_{\alpha\dot\alpha} - D
\omega^{\alpha\beta} h_\beta{}^{\dot\alpha} h_{\alpha\dot\alpha} 
\nonumber \\
 && - \frac{1}{2} \omega_\alpha{}^\beta \Psi_\beta 
e^\alpha{}_{\dot\alpha} \Psi^{\dot\alpha} + \frac{1}{2} D \Psi_\alpha
h^\alpha{}_{\dot\alpha} \Psi^{\dot\alpha} + h.c. 
\end{eqnarray}
requires the following corrections to gauge transformations:
\begin{eqnarray}
\delta h^{\alpha\dot\alpha} &=& \omega^\alpha{}_\beta 
\hat\xi^{\beta\dot\alpha} + \omega^{\dot\alpha}{}_{\dot\beta}
\hat\xi^{\alpha\dot\beta} - \hat\eta^\alpha{}_\beta
h^{\beta\dot\alpha} - \hat\eta^{\dot\alpha}{}_{\dot\beta}
h^{\alpha\dot\beta} - \frac{1}{4} (\Psi^{\dot\alpha} \zeta^\alpha +
\Psi^\alpha \zeta^{\dot\alpha}) \nonumber \\
\delta \Psi^\alpha &=& \omega^\alpha{}_\beta \zeta^\beta 
- \hat\eta^\alpha{}_\beta \Psi^\beta 
\end{eqnarray}
They are consistent with the following deformations of curvatures:
\begin{eqnarray}
\Delta T^{\alpha\dot\alpha} &=& h_\beta{}^{\dot\alpha}
\omega^{\alpha\beta} + h^\alpha{}_{\dot\beta}
\omega^{\dot\alpha\dot\beta} - \frac{1}{4} \Psi^{\dot\alpha}
\Psi^\alpha \nonumber \\
\Delta {\cal F}^\alpha &=& \omega^\alpha{}_\beta \Psi^\beta 
\end{eqnarray}
Anti-commutator gives
\begin{equation}
[ \delta_1, \delta_2] h^{\alpha\dot\alpha} = \omega^\alpha{}_\beta
\tilde\xi^{\beta\dot\alpha} + \omega^{\dot\alpha}{}_{\dot\beta}
\tilde\xi^{\alpha\dot\beta}
\end{equation}
where
\begin{equation}
\tilde\xi^{\alpha\dot\alpha} = - \frac{1}{4} (\zeta_1^\alpha
\zeta_2^{\dot\alpha} + \zeta_1^{\dot\alpha} \zeta_2^\alpha) 
\end{equation}

\section{Interactions with gravitino}

In this section, we consider the interactions of a massive spin-2
supermultiplet $(2,3/2,3/2,1)$ with a massless spin-3/2 gravitino
$\Psi^\alpha$. Supersymmetry connects only fields that differ in spin
by half, so we can treat the two superblocks $(2,3/2)$ and $(3/2,1)$
separately.

\subsection{Superblock $(2, 3/2)$}

Deformations of the unfolded equations, described in the Appendix A,
allow us to unambiguously determine the deformations of physical
fields in this superblock. For the massive spin-2 we obtain:
\begin{eqnarray}
\Delta {\cal T}^{\alpha\dot\alpha} &=& \gamma_1 (\Phi^\alpha
\Psi^{\dot\alpha} + \Phi^{\dot\alpha} \Psi^\alpha), \nonumber \\
\Delta {\cal A} &=& \frac{\epsilon}{2}\gamma_1 (\Phi^\alpha
\Psi_\alpha + \Phi^{\do\alpha} \Psi_{\dot\alpha}) -
\frac{3}{2}\gamma_1 e_{\alpha\dot\alpha} ( \phi^\alpha
\Psi^{\dot\alpha}
+ \phi^{\dot\alpha} \Psi^\alpha), \\
\Delta {\cal C} &=& - \frac{\epsilon}{2}\gamma_1 (\phi^\alpha
\Psi_\alpha + \phi^{\dot\alpha} \Psi_{\dot\alpha}). \nonumber
\end{eqnarray}
Recall that two bosons in a supermultiplet must  have opposite
parities. We chose massive spin-2 as a tensor, so parameter
$\gamma_1$ is imaginary, and parameter $\alpha_1$, bellow, is real.
On one hand,  curvature deformations, given above, determine
supertransformations for the component of massive spin-2
$(\delta \Psi^\alpha = D \zeta^\alpha)$:
\begin{eqnarray}
\delta H^{\alpha\dot\alpha} &=& \gamma_1 (\Phi^\alpha
\zeta^{\dot\alpha} + \Phi^{\dot\alpha} \zeta^\alpha), \nonumber \\
\delta A &=& \frac{\epsilon}{2}\gamma_1 (\Phi^\alpha
\zeta_\alpha + \Phi^{\do\alpha} \zeta_{\dot\alpha}) -
\frac{3}{2}\gamma_1 e_{\alpha\dot\alpha} ( \phi^\alpha
\zeta^{\dot\alpha} + \phi^{\dot\alpha} \zeta^\alpha), \\
\delta \varphi &=& \frac{\epsilon}{2}\gamma_1 (\phi^\alpha
\zeta_\alpha + \phi^{\dot\alpha} \zeta_{\dot\alpha}). \nonumber
\end{eqnarray}
On the other hand, and it will be important for what follows that they
determine corrections to the $\rho^\alpha$-transformations
($\rho^\alpha$ being the parameter of the massive spin- 3/2
$\Phi^\alpha$ gauge transformations):
\begin{equation}
\delta H^{\alpha\dot\alpha} = - \gamma_1 (\rho^\alpha
\Psi^{\dot\alpha} + \rho^{\dot\alpha} \Psi^\alpha), \qquad
\delta A = - \frac{\epsilon}{2}\gamma_1 ( \rho^\alpha
\Psi_\alpha + \rho^{\dot\alpha} \Psi_{\dot\alpha}). \label{var_b}
\end{equation}
In turn, curvature deformations for fermionic physical fields look
like:
\begin{eqnarray}
\Delta {\cal F}^\alpha &=& - \alpha_1 \Omega^{\alpha\beta} \Psi_\beta
+ \epsilon\alpha_1 e_{\beta\dot\beta} B^{\alpha\beta} \Psi^{\dot\beta}
- \epsilon M\alpha_1 H^{\alpha\dot\alpha} \Psi_{\dot\alpha} 
- \frac{M}{2}\alpha_1 A \Psi^\alpha + \frac{3\epsilon M}{2}\alpha_1
e^\alpha{}_{\dot\alpha} \varphi \Psi^{\dot\alpha}, \nonumber \\
\Delta {\cal C}^\alpha &=& \epsilon \alpha_1 \pi^{\alpha\dot\alpha}
\Psi_{\dot\alpha} + \alpha_1 B^{\alpha\beta} \Psi_\beta +
\frac{M}{2}\alpha_1 \varphi \Psi^\alpha.
\end{eqnarray}
They determine  supertransformations for the massive spin 3/2
components
\begin{eqnarray}
\delta \Phi^\alpha &=& - \alpha_1 \Omega^{\alpha\beta} \zeta_\beta
+ \epsilon\alpha_1 e_{\beta\dot\beta} B^{\alpha\beta}
\zeta^{\dot\beta} - \epsilon M\alpha_1 H^{\alpha\dot\alpha}
\zeta_{\dot\alpha} - \frac{M}{2}\alpha_1 A \Psi^\alpha +
\frac{3\epsilon M}{2}\alpha_1 e^\alpha{}_{\dot\alpha} \varphi
\zeta^{\dot\alpha}, \nonumber \\
\delta \phi^\alpha &=& - \epsilon \alpha_1 \pi^{\alpha\dot\alpha}
\zeta_{\dot\alpha} -+ \alpha_1 B^{\alpha\beta} \zeta_\beta -
\frac{M}{2}\alpha_1 \varphi \zeta^\alpha,
\end{eqnarray}
as well as corrections for the massive spin-2 gauge transformations:
\begin{equation}
\delta \Phi^\alpha = \alpha_1 \eta^{\alpha\beta} \Psi_\beta + \epsilon
M \alpha_1 \xi^{\alpha\dot\alpha} \Psi_{\dot\alpha} +
\frac{M}{2}\alpha_1 \xi \Psi^\alpha. \label{var_f}
\end{equation}

Now we consider the sum of free Lagrangians for massive spin-2 and
massive spin-3/2, and calculate their variations under local
supertransformations $\zeta^\alpha$. For the bosonic Lagrangian we
obtain
\begin{eqnarray}
\frac{1}{i} \delta {\cal L}_B &=& 4\gamma_1 \Phi_{\dot\alpha} 
e_\alpha{}^{\dot\alpha} {\cal R}^{\alpha\beta} \zeta_\beta +
4\epsilon\gamma_1 \Phi_\alpha E_{\dot\alpha(2)} 
{\cal B}^{\dot\alpha(2)} \zeta^\alpha \nonumber \\
 && - 24\gamma_1 \phi_{\dot\alpha} E_\alpha{}^{\dot\alpha}
{\cal B}^{\alpha\beta} \zeta_\beta - 12\epsilon\gamma_1 \phi_\beta
E_{\alpha\dot\alpha} \Pi^{\alpha\dot\alpha} \zeta^\beta,
\end{eqnarray}
while a variation of the fermionic Lagrangian produces
\begin{eqnarray}
\delta {\cal L}_F &=& {\cal F}_\alpha e^\alpha{}_{\dot\alpha}
[ \epsilon\alpha_1 e_{\beta\dot\beta} B^{\dot\alpha\dot\beta}
\zeta^\beta  - \epsilon M \alpha_1 H^{\beta\dot\alpha} \zeta_\beta +
\frac{3\epsilon M}{2}\alpha_1 e_\beta{}^{\dot\alpha} \varphi
\zeta^\beta ] \nonumber \\ 
 && +  {\cal F}_{\dot\alpha} e_\alpha{}^{\dot\alpha}
[ \alpha_1 \Omega^{\alpha\beta} \zeta_\beta + \frac{M}{2}\alpha_1
A \zeta^\alpha ] + 6\epsilon\alpha_1 {\cal C}_\alpha 
E^\alpha{}_{\dot\alpha} \pi^{\beta\dot\alpha} \zeta_\beta \nonumber \\
 && + 6{\cal C}_{\dot\alpha} E_\alpha{}^{\dot\alpha} 
[ \alpha_1 B^{\alpha\beta} \zeta_\beta + \frac{M}{2}\alpha_1
\varphi \zeta^\alpha ]. 
\end{eqnarray}
Using explicit expressions for the fermionic curvatures and
integrating by parts these variations can be rewritten as follows
\begin{eqnarray}
\delta {\cal L}_F &=& - \alpha_1 \Phi_{\dot\alpha} 
e_\alpha{}^{\dot\alpha} {\cal R}^{\alpha\beta}  \zeta_\beta 
- \epsilon\alpha_1 \Phi_\alpha E_{\dot\alpha(2)} 
{\cal B}^{\dot\alpha(2)} \zeta^\alpha \nonumber  \\
 && + 6\alpha_1  \phi_{\dot\alpha} E_\alpha{}^{\dot\alpha}
{\cal B}^{\alpha\beta} \zeta_\beta + 3\epsilon\alpha_1 \phi_\alpha 
E_{\beta\dot\alpha} \Pi^{\beta\dot\alpha} \zeta^\alpha \nonumber \\
 && + \alpha_1 \Phi_{\dot\alpha} e_\alpha{}^{\dot\alpha}
\Omega^{\alpha\beta} D \zeta_\beta - \epsilon\alpha_1 \Phi_\alpha
E_{\dot\alpha(2)} B^{\dot\alpha(2)} D\zeta^\alpha \nonumber \\
 && + 6\alpha_1 \phi_{\dot\alpha} E_\alpha{}^{\dot\alpha}
B^{\alpha\beta} D \zeta_\beta + 6\epsilon\alpha_1 
\phi_\alpha E^\alpha{}_{\dot\alpha} \pi^{\beta\dot\alpha} D
\zeta_\beta \nonumber \\
 && - \epsilon M\alpha_1 \Phi_\alpha e^\alpha{}_{\dot\alpha}
H^{\beta\dot\alpha} D \zeta_\beta + \frac{M}{2}\alpha_1
\Phi_{\dot\alpha} e_\alpha{}^{\dot\alpha} A D \zeta^\alpha \nonumber 
\\
 && + 3\epsilon M\alpha_1 \Phi_\alpha E^\alpha{}_\beta \varphi D
\zeta^\beta + 3M\alpha_1 \phi_{\dot\alpha} E_\alpha{}^{\dot\alpha}
\varphi D \zeta^\alpha 
\end{eqnarray}
Terms in the first two lines cancel out the bosonic variations
provided we put
$$
\alpha_1 = 4i\gamma_1
$$
Note that this relation ensures that the sum of  free
Lagrangians is invariant under global supertransformations, as it
should be. At the same time, the remaining terms can be compensated by
the (what we refer to as minimal) interactions with the gravitino:
\begin{eqnarray}
{\cal L}_1 &=& - \alpha_1 \Phi_{\dot\alpha} e_\alpha{}^{\dot\alpha}
\Omega^{\alpha\beta} \Psi_\beta + \epsilon\alpha_1 \Phi_\alpha
E_{\dot\alpha(2)} B^{\dot\alpha(2)} \Psi^\alpha \nonumber \\
 && - 6\alpha_1 \phi_{\dot\alpha} E_\alpha{}^{\dot\alpha}
B^{\alpha\beta} \Psi_\beta - 6\epsilon\alpha_1 \phi_\alpha 
E^\alpha{}_{\dot\alpha} \pi^{\beta\dot\alpha} \Psi_\beta \nonumber \\
 && + \epsilon M\alpha_1 \Phi_\alpha e^\alpha{}_{\dot\alpha}
H^{\beta\dot\alpha} \Psi_\beta - \frac{M}{2}\alpha_1 \Phi_{\dot\alpha}
e_\alpha{}^{\dot\alpha} A \Psi^\alpha \nonumber \\
 && - 3\epsilon M\alpha_1 \Phi_\alpha E^\alpha{}_\beta \varphi 
\Psi^\beta - 3M\alpha_1 \phi_{\dot\alpha} E_\alpha{}^{\dot\alpha}
\varphi \Psi^\alpha. 
\end{eqnarray}

We still have to take care on massive spin-2 and massive spin-3/2
gauge transformations. Let us consider them in turn. \\
{\bf $\rho^\alpha$-transformations} By straightforward calculations,
we obtain:
\begin{eqnarray}
\delta_\rho {\cal L}_1 &=& \alpha_1 \rho_\alpha 
e^\alpha{}_{\dot\alpha}	 {\cal R}^{\dot\alpha\dot\beta}
\Psi_{\dot\beta} - \epsilon\alpha_1 \rho_\alpha E_{\dot\alpha(2)}
B^{\dot\alpha(2)} \Psi^\alpha \nonumber \\
 && - \alpha_1 \rho_\alpha e^\alpha{}_{\dot\alpha}
\Omega^{\dot\alpha\dot\beta} D \Psi_{\dot\beta} 
- \frac{M}{2}\alpha_1 \rho_\alpha e^\alpha{}_{\dot\alpha}
A D \Psi^{\dot\alpha} - \epsilon\alpha_1 \rho_\alpha E_{\dot\alpha(2)}
B^{\dot\alpha(2)} D \Psi^\alpha \nonumber \\
 && - \epsilon M\alpha_1 \rho_\alpha e^\alpha{}_{\dot\alpha}
H^{\beta\dot\alpha} D \Psi_\beta + 3\epsilon M\alpha_1 \rho_\alpha 
E^\alpha{}_\beta \varphi D \Psi^\beta. 
\end{eqnarray}
Terms in the first line can be compensated by the corrections
\begin{equation}
\delta H^{\alpha\dot\alpha} = - \gamma_1 (\rho^\alpha
\Psi^{\dot\alpha} + \rho^{\dot\alpha} \Psi^\alpha), \qquad
\delta A = - \frac{\epsilon}{2}\gamma_1 ( \rho^\alpha
\Psi_\alpha + \rho^{\dot\alpha} \Psi_{\dot\alpha}) 
\end{equation}
in complete agreement with (\ref{var_b}). As for the remaining terms,
we need some non-minimal interactions. To completely resolve the
remaining ambiguities related to field redefinitions, we limit
ourselves to the minimum number of derivatives possible. Let us
consider
\begin{equation}
{\cal L}_2 = \kappa \Phi_\alpha H^{\alpha\dot\alpha} D
\Psi_{\dot\alpha} + h.c. 
\end{equation}
It appears that at $\kappa =  \alpha_1$ we obtain
\begin{eqnarray}
\delta ({\cal L}_1 + {\cal L}_2) &=& - \alpha_1 \rho_\alpha
\Omega^{\alpha\beta} e_\beta{}^{\dot\alpha} D \Psi_{\dot\alpha} 
- \epsilon\alpha_1 \rho_\alpha E_{\dot\alpha(2)} B^{\dot\alpha(2)} D
\Psi^\alpha \nonumber \\
 && - \frac{3M}{2}\alpha_1 \rho_\alpha e^\alpha{}_{\dot\alpha} A
D \Psi^{\dot\alpha} + 3\epsilon M\alpha_1 \rho_\alpha E^\alpha{}_\beta
\varphi D \Psi^\beta. 
\end{eqnarray}
These terms can be compensated by the following corrections
\begin{equation}
\delta \Psi^\alpha = \alpha_1 \Omega^{\alpha\beta} \rho_\beta
 + \epsilon\alpha_1 B^{\alpha\beta} e_{\beta\dot\alpha}
\rho^{\dot\alpha} + \frac{3M}{2}\alpha_1 A \rho^\alpha 
+ \frac{3\epsilon M}{2}\alpha_1 e^\alpha{}_{\dot\alpha} \varphi
\rho^{\dot\alpha}.
\end{equation}
{\bf $\eta^{\alpha(2)}$-transformations} In this case we obtain just
\begin{equation}
\delta ({\cal L}_1 + {\cal L}_2) = \alpha_1 {\cal F}_{\dot\alpha}
e_\alpha{}^{\dot\alpha} \eta^{\alpha\beta} \Psi_\beta + \alpha_1
\Phi_\alpha \eta^{\alpha\beta} e_\beta{}^{\dot\alpha}
D \Psi_{\dot\alpha}
\end{equation}
that required the following corrections
\begin{equation}
\delta \Phi^\alpha = \alpha_1 \eta^{\alpha\beta} \Psi_\beta, \qquad
\delta \Psi^\alpha = - \alpha_1 \eta^{\alpha\beta} \Phi_\beta
\end{equation}
again in agreement with (\ref{var_f}). \\
{\bf $\xi^{\alpha\dot\alpha}$-transformations} Here we obtain
\begin{equation}
\delta ({\cal L}_1 + {\cal L}_2) \approx - \epsilon M \alpha_1
{\cal F}_\alpha e^\alpha{}_{\dot\alpha} \xi^{\beta\dot\alpha}
\Psi_\beta
\end{equation}
that can be compensated (see (\ref{var_f}))
\begin{equation}
\delta \Phi^\alpha = \epsilon M \alpha_1 \xi^{\alpha\dot\alpha}
\Psi_{\dot\alpha}
\end{equation}
{\bf $\xi$-transformations} At last
\begin{equation}
\delta ({\cal L}_1 + {\cal L}_2) = \frac{M}{2}\alpha_1 
{\cal F}_{\dot\alpha} e_\alpha{}^{\dot\alpha} \xi \Psi^\alpha
+ \frac{3M}{2}\alpha_1 \Phi_{\dot\alpha} e_\alpha{}^{\dot\alpha}
\xi D \Psi^\alpha
\end{equation}
that requires
\begin{equation}
\delta \Phi^\alpha = \frac{M}{2}\alpha_1 \xi \Psi_\alpha, \qquad
\delta \Psi^\alpha = - \frac{3M}{2}\alpha_1 \xi \Phi^\alpha 
\end{equation}
Note that the corrections obtained are consistent with the following
deformation for gravitino curvature
\begin{equation}
\Delta \tilde{\cal F}^\alpha = \alpha_1 \Omega^{\alpha\beta}
\Phi_\beta  + \epsilon\alpha_1 B^{\alpha\beta} e_{\beta\dot\alpha}
\Phi^{\dot\alpha} + \frac{3M}{2}\alpha_1 A \Phi^\alpha 
- 3\epsilon\alpha_1 e^\alpha{}_{\dot\alpha} \varphi \Phi^{\dot\alpha}
\end{equation}
{\bf Algebraic structure} All fields we deal with are either gauge
fields or Stueckelberg fields which have non-homogeneous gauge
transformations. Thus, having in our disposal corrections to the gauge
transformations linear in fields, we can already calculate
(anti)commutators to the zeroth order, which reveal  important
information about the algebraic structure underlying our model. By
calculating (anti)commutators we obtain for massive spin-2
\begin{eqnarray}
\ [\delta_1, \delta_2] H^{\alpha\dot\alpha} &=& D
\tilde\xi^{\alpha\dot\alpha} + e_\beta{}^{\dot\alpha}
\tilde\eta^{\alpha\beta} + e^\alpha{}_{\dot\beta}
\tilde\eta^{\dot\alpha\dot\beta} + M e^{\alpha\dot\alpha}
\tilde\xi, \nonumber \\
\ [\delta_1, \delta_2] A &=& D \tilde\xi + M e_{\alpha\dot\alpha}
\tilde\xi^{\alpha\dot\alpha}, \\
\ [\delta_1, \delta_2] \varphi &=& - M \tilde\xi, \nonumber
\end{eqnarray}
where
\begin{equation}
\tilde\xi^{\alpha\dot\alpha} = \gamma_1 (\rho^\alpha
\zeta^{\dot\alpha} + \rho^{\dot\alpha} \zeta^\alpha), \qquad
\tilde\eta^{\alpha(2)} = \frac{\epsilon M}{2}\gamma_1
\rho^\alpha \zeta^\alpha, \qquad \tilde\xi = 
\frac{\epsilon M}{2}\gamma_1 (\rho^\alpha \zeta_\alpha +
\rho^{\dot\alpha} \zeta_{\dot\alpha}). 
\end{equation}
At the same time, for the massive spin 3/2 we obtain
\begin{equation}
[\delta_1, \delta_2] \Phi^\alpha = D \tilde\rho^\alpha + \epsilon M
e^\alpha{}_{\dot\alpha} \tilde\rho^{\dot\alpha}, \qquad
[\delta_1, \delta_2] \phi^\alpha = - M \tilde\rho^\alpha,
\end{equation}
where
\begin{equation}
\tilde\rho^\alpha = - \alpha_1 \eta^{\alpha\beta} \zeta_\beta
- \epsilon M \alpha_1 \xi^{\alpha\dot\alpha} \zeta_{\dot\alpha}
- \frac{M}{2} \xi \zeta^\alpha. 
\end{equation}

\subsection{Superblock $(3/2, 1)$}

Using the deformation procedure for the unfolded equations described
in Appendix B, we obtain the following deformations for massive spin-1
curvatures:
\begin{eqnarray}
\Delta \tilde{\cal A} &=& 2\gamma_1 (\Phi^\alpha \Psi_\alpha -
\Phi^{\dot\alpha} \Psi_{\dot\alpha}) + 2\epsilon\gamma_1
e_{\alpha\dot\alpha} ( \phi^\alpha \Psi^{\dot\alpha} -
\phi^{\dot\alpha} \psi^\alpha), \nonumber \\
\Delta \tilde{\cal C} &=& - \gamma_1 ( \phi^\alpha \Psi_\alpha
- \phi^{\dot\alpha} \Psi_{\dot\alpha}. 
\end{eqnarray}
Note that our choice that massive spin-2 is a tensor implies that
massive spin-1 is a pseudo-vector, so the parameter $\gamma_1$
is real and the parameter $\alpha_1$, below, is imaginary. In turn,
these deformations determine both supertransformations
\begin{eqnarray}
\delta \tilde{A} &=& 2\gamma_1 (\Phi^\alpha \zeta_\alpha -
\Phi^{\dot\alpha} \zeta_{\dot\alpha}) + 2\epsilon\gamma_1
e_{\alpha\dot\alpha} (\phi^\alpha \zeta^{\dot\alpha} -
\phi^{\dot\alpha} \zeta^\alpha), \nonumber \\
\delta \tilde\varphi &=& \gamma_1 (\phi^\alpha \zeta_\alpha -
\phi^{\dot\alpha} \zeta_{\dot\alpha}), 
\end{eqnarray}
as well as corrections to the massive spin 3/2 
$\rho^\alpha$-transformations
\begin{equation}
\delta \tilde{A} = - 2\gamma_1 ( \rho^\alpha \Psi_\alpha -
\rho^{\dot\alpha} \Psi_{\dot\alpha}). \label{var_3}
\end{equation}
Similarly, for  the massive spin 3/2 curvatures deformations we obtain
\begin{eqnarray}
\Delta {\cal F}^\alpha &=& \alpha_1 e_{\beta\dot\beta}
\tilde{B}^{\alpha\beta} \Psi^{\dot\beta} + \frac{\epsilon
M}{2}\alpha_1 \tilde{A} \Psi^\alpha  + M\alpha_1
e^\alpha{}_{\dot\alpha} \tilde\varphi \Psi^{\dot\alpha}, \nonumber \\
\Delta {\cal C}^\alpha &=& \frac{2}{3}\alpha_1
\tilde\pi^{\alpha\dot\alpha} \Psi_{\dot\alpha} - \frac{\epsilon}{3}
\alpha_1 \tilde{B}^{\alpha\beta} \Psi_\beta - \epsilon M\alpha_1
\tilde\varphi \Psi^\alpha,
\end{eqnarray}
which correspond to the following supertransformations
\begin{eqnarray}
\delta \Phi^\alpha &=& \alpha_1 e_{\beta\dot\beta} 
\tilde{B}^{\alpha\beta} \zeta^{\dot\beta} + \frac{\epsilon
M}{2}\alpha_1 \tilde{A} \zeta^\alpha + M\alpha_1
e^\alpha{}_{\dot\alpha} \tilde\varphi \zeta^{\dot\alpha}, \nonumber \\
\delta \phi^\alpha &=& - \frac{2}{3}\alpha_1
\tilde\pi^{\alpha\dot\alpha} \zeta_{\dot\alpha} + \frac{\epsilon}{3}
\alpha_1 \tilde{B}^{\alpha\beta} \zeta_\beta + \epsilon M\alpha_1
\tilde\varphi \zeta^\alpha, 
\end{eqnarray}
as well as corrections for the massive spin 1 gauge transformations
\begin{equation}
\delta \Phi^\alpha = - \frac{\epsilon M}{2}\alpha_1 \tilde\xi
\Psi^\alpha. \label{var_4}
\end{equation}

Now we consider the sum of the free Lagrangians for massive spin-3/2
and massive spin-1 and calculate their variation under local
supertransformations. For the variation of massive spin-1 we obtain
\begin{equation}
\frac{1}{i} \delta {\cal L}_B =  4\gamma_1 \Phi_\alpha
E_{\dot\alpha(2)} \tilde{\cal B}^{\dot\alpha(2)}
\zeta^\alpha - 8\epsilon\gamma_1 \phi_{\dot\alpha}
\tilde{\cal B}^{\dot\alpha\dot\beta} E^\alpha{}_{\dot\beta}
\zeta_\alpha - 8\gamma_1 \phi_\alpha E_{\beta\dot\alpha}
\tilde\Pi^{\beta\dot\alpha} \zeta^\alpha
\end{equation}
while variation of the massive spin-3/2 produces:
\begin{eqnarray}
\delta {\cal L}_F &=& {\cal F}_\alpha e^\alpha{}_{\dot\alpha}
[ - \alpha_1 e_{\beta\dot\beta} \tilde{B}^{\dot\alpha\dot\beta}
\zeta^\beta - M\alpha_1 e_\beta{}^{\dot\alpha} \tilde\varphi
\zeta^\beta ] - {\cal F}_{\dot\alpha} e_\alpha{}^{\dot\alpha}
\frac{\epsilon M}{2} \alpha_1 \tilde{A} \zeta^\alpha \nonumber \\
 && - 6 {\cal C}_{\dot\alpha} E_\alpha{}^{\dot\alpha} 
[ \frac{\epsilon}{3} \alpha_1 \tilde{B}^{\alpha\beta} \zeta_\beta +
\epsilon M\alpha_1 \tilde\varphi \zeta^\alpha] - 4 {\cal C}_\alpha
E^\alpha{}_{\dot\alpha} \alpha_1 \tilde\pi^{\beta\dot\alpha}
\zeta_\beta.
\end{eqnarray}
Using explicit expressions for fermionic curvatures and integrating
by parts these variations can be rewritten as follows:
\begin{eqnarray}
\delta {\cal L}_F &=& \alpha_1 \Phi_\alpha E_{\dot\alpha(2)} 
\tilde{\cal B}^{\dot\alpha(2)} \zeta^\alpha  - 2\epsilon\alpha_1
\phi_{\dot\alpha} E_\alpha{}^{\dot\alpha} \tilde{\cal B}^{\alpha\beta}
\zeta_\beta - 4\alpha_1 \phi_\alpha E^\alpha{}_{\dot\alpha}
\tilde\Pi^{\beta\dot\alpha} \zeta_\beta \nonumber \\
 && + \alpha_1 \Phi_\alpha E_{\dot\alpha(2)} \tilde{B}^{\dot\alpha(2)}
D \zeta^\alpha - \frac{\epsilon M}{2}\alpha_1 \Phi_{\dot\alpha}
e_\alpha{}^{\dot\alpha} \tilde{A} D \zeta^\alpha - 2M\alpha_1
\Phi_\alpha E^\alpha{}_\beta \tilde\varphi D \zeta^\beta \nonumber \\
 && - 2\epsilon \alpha_1 \phi_{\dot\alpha} E_\alpha{}^{\dot\alpha}
\tilde{B}^{\alpha\beta} D \zeta_\beta - 4\alpha_1 \phi_\alpha
E^\alpha{}_{\dot\alpha} \tilde\pi^{\beta\dot\alpha} D \zeta_\beta 
- 6\epsilon M\alpha_1 \phi_{\dot\alpha} E_\alpha{}^{\dot\alpha}
\tilde\varphi D \zeta^\alpha
\end{eqnarray}
At $\alpha_1 = - 4i\gamma_1$, the terms in the first line cancel out
bosonic variations, so that the sum of the free Lagrangians is
invariant under global supertransformations. The remaining terms
require minimal interactions:
\begin{eqnarray}
{\cal L}_1 &=& - \alpha_1 \Phi_\alpha E_{\dot\alpha(2)}
\tilde{B}^{\dot\alpha(2)} \Psi_\alpha + \frac{\epsilon M}{2}\alpha_1
\Phi_{\dot\alpha} e_\alpha{}^{\dot\alpha} \tilde{A} \Psi^\alpha +
2M\alpha_1 \Phi_\alpha E^\alpha{}_\beta \tilde\varphi \Psi^\beta
\nonumber \\
 && + 2\epsilon \alpha_1 \phi_{\dot\alpha} E_\alpha{}^{\dot\alpha}
\tilde{B}^{\alpha\beta} \Psi_\beta + 4\alpha_1 \phi_\alpha
E^\alpha{}_{\dot\alpha} \tilde\pi^{\beta\dot\alpha} \Psi_\beta 
+ 6\epsilon M\alpha_1 \phi_{\dot\alpha} E_\alpha{}^{\dot\alpha}
\tilde\varphi \Psi^\alpha. 
\end{eqnarray}
We still have to take care on the massive spin 1 and massive spin 3/2
gauge transformations. \\
{\bf $\tilde\xi$-transformations} They produce simply
\begin{equation}
\delta {\cal L}_1 = - \frac{\epsilon M}{2}\alpha_1 D \Psi_\alpha
e^\alpha{}_{\dot\alpha} \Phi^{\dot\alpha} \tilde\xi - 
\frac{\epsilon M}{2}\alpha_1 {\cal F}_{\dot\alpha} 
e_\alpha{}^{\dot\alpha} \Psi^\alpha \tilde\xi
\end{equation}
and these variations can be compensated by he following corrections
\begin{equation}
\delta \Psi^\alpha = - \frac{\epsilon M}{2}\alpha_1 \Phi^\alpha
\tilde\xi, \qquad \delta \Phi^\alpha = \frac{\epsilon M}{2}\alpha_1
\Psi^\alpha \tilde\xi
\end{equation}
in agreement with (\ref{var_4}). \\
{\bf $\rho^\alpha$-transformations} In this case we obtain
\begin{eqnarray}
\delta {\cal L}_1 &=& - D \Psi_\alpha e^\alpha{}_{\dot\alpha}
e_{\beta\dot\beta} \tilde{B}^{\dot\alpha\dot\beta} \rho^\beta +
\frac{\epsilon M}{2} D \Psi_{\dot\alpha} e_\alpha{}^{\dot\alpha}
\tilde{A} \rho^\alpha + M D\Psi_\alpha e^\alpha{}_{\dot\alpha}
e_\beta{}^{\dot\alpha} \tilde\varphi \rho^\beta \nonumber \\
 && + \frac{1}{2} [ E_{\beta(2)} \tilde{\cal B}^{\beta(2)} -
E_{\dot\alpha(2)} \tilde{\cal B}^{\dot\alpha(2)}] \rho^\alpha
\Psi_\alpha,
\end{eqnarray}
what can be compensated by the following corrections
\begin{eqnarray}
\delta \Psi^\alpha &=& - \alpha_1 e_{\beta\dot\alpha}
\tilde{B}^{\alpha\beta} \rho^{\dot\alpha} + \frac{\epsilon M}{2}
\alpha_1 \tilde{A} \rho^\alpha + M\alpha_1 e^\alpha{}_{\dot\alpha}
\tilde\varphi \rho^{\dot\alpha}, \nonumber \\
\delta \tilde{A} &=& - \frac{i}{2}\alpha_1 \rho^\alpha \Psi_\alpha +
h.c. 
\end{eqnarray}
again in agreement with (\ref{var_3}).

{\bf Algebraic structure} Calculating the anti-commutators of 
$\zeta^\alpha$ and $\rho^\alpha$ transformations we obtain
\begin{equation}
[\delta_1, \delta_2] \tilde{A} =  D \hat\xi, \qquad
[\delta_1, \delta_2 ] \tilde\varphi = - \frac{M}{2} \hat\xi,
\end{equation}
where 
\begin{equation}
\hat\xi = 2\gamma_1 (\rho^\alpha \zeta_\alpha - \rho^{\dot\alpha}
\zeta_{\dot\alpha}). 
\end{equation}
Similarly, commutators of massive spin 1 gauge transformations and
supertransformations produce
\begin{equation}
[\delta_1, \delta_2] \Phi^\alpha = ( D \hat\rho^\alpha + \epsilon M
e^\alpha{}_{\dot\alpha} \hat\rho^{\dot\alpha}), \qquad
[\delta_1, \delta_2  ] \phi^\alpha = - M \hat\rho^\alpha,  
\end{equation}
where
\begin{equation}
\hat\rho^\alpha = \frac{\epsilon M}{2}\alpha_1 \tilde\xi \zeta^\alpha.
\end{equation}

\section{Interactions with graviton}

There are well known substitution rules to construct minimal
gravitational interactions which in our formalism look like:
\begin{equation}
e^{\alpha\dot\alpha} \Rightarrow e^{\alpha\dot\alpha} +
h^{\alpha\dot\alpha}, \qquad D \phi^\alpha \Rightarrow
D \phi^\alpha + \omega^\alpha{}_\beta \phi^\beta.
\end{equation}
For the massive spin 2 they produce the following deformations for
curvatures:
\begin{eqnarray}
\Delta {\cal T}^{\alpha\dot\alpha} &=& \omega^\alpha{}_\beta
H^{\beta\dot\alpha} + \omega^{\dot\alpha}{}_{\dot\beta}
H^{\alpha\dot\beta} + h_\beta{}^{\dot\alpha} \Omega^{\alpha\beta}
+ h^\alpha{}_{\dot\beta} \Omega^{\dot\alpha\dot\beta} + M
h^{\alpha\dot\alpha} A, \nonumber \\
\Delta {\cal A} &=& e^\alpha{}_{\dot\alpha} h^{\alpha\dot\alpha}
B_{\alpha(2)} + e_\alpha{}^{\dot\alpha} h^{\alpha\dot\alpha}
B_{\dot\alpha(2)}  + M h_{\alpha\dot\alpha} H^{\alpha\dot\alpha}, \\
\Delta {\cal C} &=& h_{\alpha\dot\alpha} \pi^{\alpha\dot\alpha}.
\nonumber 
\end{eqnarray}
This in turn leads to the following corrections to the gauge
transformations:
\begin{eqnarray}
\delta_1 H^{\alpha\dot\alpha} &=& \eta^{\alpha\beta} 
h_\beta{}^{\dot\alpha} + \eta^{\dot\alpha\dot\beta}
h^\alpha{}_{\dot\beta} + \omega^\alpha{}_\beta \xi^{\beta\dot\alpha}
+ \omega^{\dot\alpha}{}_{\dot\beta} \xi^\alpha{}_{\dot\beta}
+ M h^{\alpha\dot\alpha} \xi \nonumber \\
\delta_1 A &=& M h^{\alpha\dot\alpha} \xi_{\alpha\dot\alpha}. 
\end{eqnarray}
{\bf $\eta^{\alpha(2)}$-transformations} Calculating a total variation
we obtain:
\begin{equation}
\delta_0 {\cal L}_1 + \delta_1 {\cal L}_0 = 2 ( D
\omega^{\alpha\gamma} \eta_\gamma{}^\beta + D \omega^{\beta\gamma}
\eta_\gamma{}^\alpha) e_\beta{}^{\dot\alpha} H_{\alpha\dot\alpha}.
\end{equation}
Let us introduce non-minimal interaction
\begin{equation}
{\cal L}_2 = \kappa D \omega^{\alpha\beta} H_\alpha{}^{\dot\alpha}
H_{\beta\dot\alpha} + h.c.
\end{equation}
This leads to
\begin{equation}
\delta_0 ({\cal L}_1 +  {\cal L}_2) + \delta_1 {\cal L}_0 =
 2 [D \omega^{\alpha\beta} e_\beta{}^{\dot\alpha} 
 + a_0 D \omega^{\dot\alpha\dot\beta} e^\alpha{}_{\dot\beta} ]
\eta_\alpha{}^\gamma H_{\gamma\dot\alpha} + 2(1+\kappa) D
\omega^{\alpha\beta} e^\gamma{}_{\dot\alpha} H_\alpha{}^{\dot\alpha}
\eta_{\beta\gamma}.
\end{equation}
We put $\kappa = - 1$, then the remaining terms can be compensated by
correction
\begin{equation}
\delta h^{\alpha\dot\alpha} = \eta^{\alpha\beta} 
H_\beta{}^{\dot\alpha}.
\end{equation}
{\bf $\xi^{\alpha\dot\alpha}$-transformations} Here we obtain
\begin{equation}
\delta_0 ({\cal L}_1 +  {\cal L}_2) + \delta_1 {\cal L}_0 =
- 2[ D \omega^{\alpha\beta} e_\beta{}^{\dot\alpha} +
D \omega^{\dot\alpha\dot\beta} e^\alpha{}_{\dot\beta}]
(\Omega_\alpha{}^{\gamma} \xi_{\gamma\dot\alpha} - M A
\xi_{\alpha\dot\alpha}),
\end{equation}
that can be compensated by the corrections
\begin{equation}
\delta h^{\alpha\dot\alpha} = - \Omega^{\alpha\beta}
\xi_\beta{}^{\dot\alpha} + M A \xi^{\alpha\dot\alpha}.
\end{equation}
{\bf $\xi$-transformations} They produce
\begin{equation}
\delta_0 ({\cal L}_1 +  {\cal L}_2) + \delta_1 {\cal L}_0 =
- 2M [ D \omega^{\alpha\beta} e_\beta{}^{\dot\alpha} +
D \omega^{\dot\alpha\dot\beta} e^\alpha{}_{\dot\beta}]
H_{\alpha\dot\alpha} \xi
\end{equation}
and this requires
\begin{equation}
\delta h^{\alpha\dot\alpha} = - M H^{\alpha\dot\alpha} \xi.
\end{equation}
Collecting all corrections we find that they correspond to the
following deformation to the torsion
\begin{equation}
\Delta T^{\alpha\dot\alpha} = \Omega^\alpha{}_\beta
H^{\beta\dot\alpha} + M A H^{\alpha\dot\alpha}.
\end{equation}

For the massive spin 3/2 the substitution rules produce the following
corrections to the curvatures:
\begin{eqnarray}
\Delta {\cal F}^\alpha &=& \omega^\alpha{}_\beta \Phi^\beta + \epsilon
M h^\alpha{}_{\dot\alpha} \Phi^{\dot\alpha} - M 
(e^\alpha{}_{\dot\alpha} h^{\beta\dot\alpha} + e^\beta{}_{\dot\alpha}
h^{\alpha\dot\alpha}) \phi_\beta, \nonumber \\
\Delta {\cal C}^\alpha &=& \omega^\alpha{}_\beta \phi^\beta + \epsilon
M h^\alpha{}_{\dot\alpha} \phi^{\dot\alpha}, 
\end{eqnarray}
which give the following correction for the 
$\rho^\alpha$-transformations:
\begin{equation}
\delta_1 \Phi^\alpha = \omega^\alpha{}_\beta \rho^\beta + \epsilon M
h^\alpha{}_{\dot\alpha} \rho^{\dot\alpha}. 
\end{equation}
Calculating the variations we obtain
\begin{equation}
\delta_0 {\cal L}_1 + \delta_1 {\cal L}_0 = \frac{1}{2}
[ D \omega^{\alpha\beta} e_\beta{}^{\dot\alpha} +
D \omega^{\dot\alpha\dot\beta} e^\alpha{}_{\dot\beta}]
\Phi_{\dot\alpha} \rho_\alpha
\end{equation}
They can be compensated by the following corrections
\begin{equation}
\delta_1 h^{\alpha\dot\alpha} = \frac{1}{4} \Phi^{\dot\alpha}
\rho^\alpha + h.c.
\end{equation}
and this in turn corresponds  to
\begin{equation}
\Delta T^{\alpha\dot\alpha} = \frac{1}{4} \Phi^\alpha
\Phi^{\dot\alpha}. 
\end{equation}

As is well known, for spin 1 standard substitution rules produce gauge
invariant Lagrangian without any need to introduce non-minimal
interactions or corrections to the gauge transformations.

\section{Supermultiplet}

Recall that our coordinate-free description of Minkowski space is
based on the background frame $e^{\alpha\dot\alpha}$ and a
background Lorentz connection entering into the covariant derivative
$D$. As a result, the explicit form of the gauge invariant curvatures
determines global translations and Lorentz transformations that leave
our free Lagrangians invariant. The Lorentz transformations have a
universal form, e.g.
\begin{equation}
\delta H^{\alpha\dot\alpha} = - \hat\eta^\alpha{}_\beta
H^{\beta\dot\alpha} - \hat\eta^{\dot\alpha}{}_{\dot\beta}
H^{\alpha\dot\beta}.
\end{equation}
As for the global translations, from the massive spin 2 curvatures
(\ref{cur_2}) we directly obtain
\begin{eqnarray}
\delta H^{\alpha\dot\alpha} &=&  \Omega^\alpha{}_\beta 
\hat\xi^{\beta\dot\alpha} + \Omega^{\dot\alpha}{}_{\dot\beta}
\hat\xi^{\alpha\dot\beta} - M A \hat\xi^{\alpha\dot\alpha}, \nonumber
 \\
\delta A &=& 2 ( e^\alpha{}_{\dot\alpha} \hat\xi^{\alpha\dot\alpha}
B_{\alpha(2)} + e_\alpha{}^{\dot\alpha} \hat\xi^{\alpha\dot\alpha}
B_{\dot\alpha(2)}) - M H^{\alpha\dot\alpha}
\hat\xi_{\alpha\dot\alpha}, \label{com_2} \\
\delta \varphi &=& - \pi^{\alpha\dot\alpha}
\hat\xi_{\alpha\dot\alpha}. \nonumber
\end{eqnarray}
Similarly, from (\ref{cur_1}) it follows that
\begin{eqnarray}
\delta \tilde{A} &=& 2 ( e^\alpha{}_{\dot\alpha}
\hat\xi^{\alpha\dot\alpha} \tilde{B}_{\alpha(2)} + 
e_\alpha{}^{\dot\alpha} \hat\xi^{\alpha\dot\alpha} 
\tilde{B}_{\dot\alpha(2)}), \nonumber \\
\delta \tilde\varphi &=& - \tilde\pi^{\alpha\dot\alpha}
\hat\xi_{\alpha\dot\alpha}. \label{com_1}
\end{eqnarray}
It is a straightforward task to check that these transformations
correspond to the standard Poincare algebra.

Now let us consider all four superblocks with four a priory
independent coupling constants: 
$$
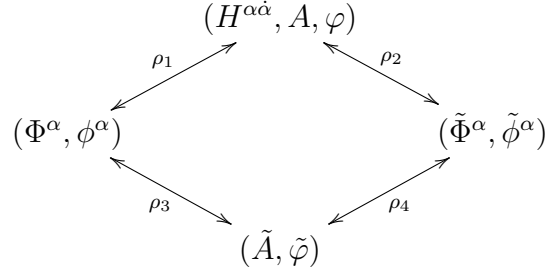
\xymatrix{
  &  (H^{\alpha\dot\alpha}, A, \varphi) \ar@{<->}[dr]^-{\rho_2}  &  \\
 (\Phi^\alpha, \phi^\alpha) \ar@{<->}[ur]^-{\rho_1} & &
(\tilde\Phi^\alpha, \tilde\phi^\alpha) \ar@{<->}[dl]^-{\rho_4} \\
  & (\tilde{A}, \tilde\varphi)  \ar@{<->}[ul]^-{\rho_3}  &  }
$$
For the supertransformations for massive spin-2 we have
\begin{eqnarray}
\delta H^{\alpha\dot\alpha} &=& - \frac{i}{4}\rho_1 (\Phi^\alpha
\zeta^{\dot\alpha} + \Phi^{\dot\alpha} \zeta^\alpha) -
\frac{i}{4}\rho_2 (\tilde\Phi^\alpha \zeta^{\dot\alpha} +
\tilde\Phi^{\dot\alpha} \zeta^\alpha), \nonumber \\
\delta A &=& - \frac{i}{8}\rho_1 (\Phi^\alpha \zeta_\alpha +
\Phi^{\dot\alpha} \zeta_{\dot\alpha}) + \frac{3i}{8}\rho_1
e_{\alpha\dot\alpha} (\phi^\alpha \zeta^{\dot\alpha} +
\phi^{\dot\alpha} \zeta^\alpha) \nonumber \\
 && + \frac{i}{8}\rho_2 (\tilde\Phi^\alpha \zeta_\alpha +
\tilde\Phi^{\dot\alpha} \zeta_{\dot\alpha}) + \frac{3i}{8}\rho_2
e_{\alpha\dot\alpha} (\tilde\phi^\alpha \zeta^{\dot\alpha} + 
\tilde\phi^{\dot\alpha} \zeta^\alpha), \\
\delta \varphi &=& - \frac{i}{8}\rho_1 (\phi^\alpha \zeta_\alpha +
\phi^{\dot\alpha} \zeta_{\dot\alpha}) + \frac{i}{8}\rho_2
(\tilde\phi^\alpha \zeta_\alpha + \tilde\phi^{\dot\alpha}
\zeta_{\dot\alpha}). \nonumber
\end{eqnarray}
Similarly, for the supertransformations for massive spin-3/2 we obtain
\begin{eqnarray}
\delta \Phi^\alpha &=& - \rho_1 \Omega^{\alpha\beta} \zeta_\beta
+ \rho_1 e_{\beta\dot\beta} B^{\alpha\beta}
\zeta^{\dot\beta} -  M\rho_1 H^{\alpha\dot\alpha}
\zeta_{\dot\alpha} - \frac{M}{2}\rho_1 A \Psi^\alpha +
\frac{3M}{2}\rho_1 e^\alpha{}_{\dot\alpha} \varphi
\zeta^{\dot\alpha} \nonumber \\
 && + i\rho_3 e_{\beta\dot\alpha} \tilde{B}^{\alpha\beta}
\zeta^{\dot\alpha} + \frac{iM}{2}\rho_3 \tilde{A} \zeta^\alpha +
iM\rho_3 e^\alpha{}_{\dot\alpha} \tilde\varphi \zeta^{\dot\alpha},
\nonumber \\
\delta \phi^\alpha &=& - \rho_1 \pi^{\alpha\dot\alpha}
\zeta_{\dot\alpha} - \rho_1 B^{\alpha\beta} \zeta_\beta -
\frac{M}{2}\rho_1 \varphi \zeta^\alpha \\
 && - \frac{2i}{3}\rho_3 \tilde\pi^{\alpha\dot\alpha} 
\zeta_{\dot\alpha} + \frac{i}{3}\rho_3 \tilde{B}^{\alpha\beta}
\zeta_\beta + iM\rho_3 \tilde\varphi \zeta^\alpha, \nonumber 
\end{eqnarray}
\begin{eqnarray}
\delta \tilde\Phi^\alpha &=& - \rho_2 \Omega^{\alpha\beta} \zeta_\beta
- \rho_2 e_{\beta\dot\beta} B^{\alpha\beta}
\zeta^{\dot\beta} + M\rho_2 H^{\alpha\dot\alpha}
\zeta_{\dot\alpha} - \frac{M}{2}\rho_2 A \Psi^\alpha -
\frac{3M}{2}\rho_2 e^\alpha{}_{\dot\alpha} \varphi
\zeta^{\dot\alpha} \nonumber \\
 && - i\rho_4 e_{\beta\dot\alpha} \tilde{B}^{\alpha\beta}
\zeta^{\dot\alpha} + \frac{iM}{2}\rho_4 A \zeta^\alpha - iM\rho_4
e^\alpha{}_{\dot\alpha} \tilde\varphi \zeta^{\dot\alpha}, \nonumber \\
\delta \tilde\phi^\alpha &=& \rho_2 \pi^{\alpha\dot\alpha}
\zeta_{\dot\alpha} - \rho_2 B^{\alpha\beta} \zeta_\beta -
\frac{M}{2}\rho_2 \varphi \zeta^\alpha \\
 && + \frac{2i}{3}\rho_4 \pi^{\alpha\dot\alpha} \zeta_{\dot\alpha}
 + \frac{i}{3}\rho_4 \tilde{B}^{\alpha\beta} \zeta_\beta 
 + iM\rho_4 \tilde\varphi \zeta^\alpha. \nonumber 
\end{eqnarray}
At last,  the supertransformations for massive spin-1 take the form:
\begin{eqnarray}
\delta \tilde{A} &=& - \frac{1}{2}\rho_3 (\Phi^\alpha \zeta_\alpha -
\Phi^{\dot\alpha} \zeta_{\dot\alpha}) - \frac{1}{2}\rho_3 
e_{\alpha\dot\alpha} ( \phi^\alpha \zeta^{\dot\alpha} -
\phi^{\dot\alpha} \zeta^\alpha) \nonumber \\
 && + \frac{1}{2}\rho_4 (\tilde\Phi^\alpha \zeta_\alpha -
\tilde\Phi^{\dot\alpha} \zeta_{\dot\alpha}) - \frac{1}{2}\rho_4 
e_{\alpha\dot\alpha} ( \tilde\phi^\alpha \zeta^{\dot\alpha} -
\tilde\phi^{\dot\alpha} \zeta^\alpha), \\
\delta \tilde\varphi &=& - \frac{1}{4}\rho_3 (\phi^\alpha
\zeta_\alpha - \phi^{\dot\alpha} \zeta_{\dot\alpha}) +
\frac{1}{4}\rho_4 (\tilde\phi^\alpha \zeta_\alpha -
\tilde\phi^{\dot\alpha} \zeta_{\dot\alpha}). \nonumber
\end{eqnarray}

Now calculating the anti-commutators of the two supertransformations
on the bosonic components, we find that closure of the superalgebra
requires that four coupling constants must satisfy the following
relations (recall that the gravitational coupling constant is set to
1):
\begin{equation}
\rho_1{}^2 = \rho_2{}^2 = \frac{1}{2}, \qquad
\rho_3{}^2 = \rho_4{}^2 = \frac{3}{4}, \qquad
\rho_1 \rho_3 = \rho_2 \rho_4
\end{equation}
In this, anti-commutators correctly reproduce (\ref{com_2}) and
(\ref{com_1}) where
\begin{equation}
\hat\xi^{\alpha\dot\alpha} = - \frac{i}{4} (\zeta_1^\alpha
\zeta_2^{\dot\alpha} + \zeta_1^{\dot\alpha} \zeta_2^\alpha)
\end{equation}

\section{Conclusion}

In this work, we consider the interaction of the massless $N=1$
supergravity with a massive $(2,3/2,3/2,1)$ supermultiplet as a
possible candidate for a supersymmetric extension to bigravity. We use
a gauge invariant description for massive spin-2, spin-3/2 and spin-1
fields. As a result of this, ambiguities arise related to possible
field redefinitions, which are fixed by using a recently proposed
method \cite{Zin24} based on unfolded equations. Additionally,
interactions with minimum possible number of derivatives (two
derivatives for bosonic vertices and one derivative for fermionic
ones) are considered. This completely fixes the entire construction.

The absence of higher-order derivatives means that our model admits a
non-singular massless limit. It is well known that for theories with
multiple massless spin-2 fields, there are no-go theorems
\cite{BDGH00} that prohibit any non-trivial interactions between them.
It is instructive to examine what happens in bigravity in the massless
limit. At the cubic level, bigravity can be represented schematically
\cite{Zin12,Zin13} as ($h$ -- massless, $H$ - massive):
$$
{\cal L}_1 = g hhh \quad (\oplus \quad hhH) \quad \oplus \quad g hHH
\quad \oplus \quad \tilde{g} HHH
$$
Here, the first term represents the usual self-interaction of massless
spin-2, while the second term, "two-massless, one massive", is absent
because it would require four derivatives. The third term describes
the gravitational interaction of a massive spin-2 with  the same
coupling constant as in the first one; and the last term with a priory
independent coupling constant corresponds to the massive spin-2
selfinteraction. This particular structure (namely the absence of a
second term and equality of coupling constants in the first and third
terms) that allows us by a simple field redefinition
\begin{eqnarray*}
\tilde{h} &=& \cos(\theta)  h + \sin(\theta) H \\
\tilde{H} &=& - \sin(\theta) h + \cos(\theta) H
\end{eqnarray*}
where
$$
\tan(2\theta) = \frac{2g}{\tilde{g}}
$$
to completely separate a massless limit into two  independent
components in agreement with the no-go theorems.

But similar no-go results also hold for the massless multi 
supergravities as well \cite{BHN02}. So it would be interesting and
important to analyse  what happens in the supersymmetric case. To do
this, we have to add self-interactions for the massive spin-2
supermultiplet (see e.g.  \cite{Mal13a,Zin18,EJP22}), but we leave
this task for the future.

\appendix

\section{Superblock $(2,3/2)$}

Here we present our results on the deformation of the unfolded
equations in the presence of the massless gravitino. Naturally,
in order for such a deformation to be possible, each field must have a
superpartner. Due to the fact that supersymmetry directly connects
only fields that differ in spin by half, we can consider two
superblocks $(2,3/2)$ and $(3/2,1)$ separately.

\subsection{Deformation for massive spin 2}

The most general ansatz for the unfolded equations of the gauge
invariant zero-forms looks as follows:
\begin{eqnarray}
0 &=& D W^{\alpha(4+k)\dot\alpha(k)} - e_{\beta\dot\beta}
W^{\alpha(4+k)\beta\dot\alpha(k)\dot\beta} + a_{1,k}
e^\alpha{}_{\dot\beta} B^{\alpha(3+k)\dot\alpha(k)\dot\beta}
+ b_{1,k} e^{\alpha\dot\alpha} W^{\alpha(3+k)\dot\alpha(k-1)}
\nonumber \\
 && + \gamma_{1,k} Y^{\alpha(4+k)\dot\alpha(k)\dot\beta}
\Psi_{\dot\beta} + \delta_{1,k} Y^{\alpha(3+k)\dot\alpha(k)}
\Psi^\alpha, \\
0 &=&  B^{\alpha(3+k)\dot\alpha(k+1)} - e_{\beta\dot\beta}
B^{\alpha(3+k)\beta\dot\alpha(k+1)\dot\beta} + a_{2,k}
e_\beta{}^{\dot\alpha} W^{\alpha(3+k)\beta\dot\alpha(k)} \nonumber \\
 && + a_{3,k} e^\alpha{}_{\dot\beta} 
\pi^{\alpha(2+k)\dot\alpha(k+1)\dot\beta} + b_{2,k}
e^{\alpha\dot\alpha} B^{\alpha(2+k)\dot\alpha(k)} \nonumber \\
  && + \gamma_{2,k} Y^{\alpha(3+k)\beta\dot\alpha(k+1)} \Psi_\beta
 + \delta_{2,k} Y^{\alpha(3+k)\dot\alpha(k)} \Psi^{\dot\alpha}
\nonumber \\
 && + \gamma_{3,k} \phi^{\alpha(3+k)\dot\alpha(k+1)\dot\beta}
\Psi_{\dot\beta} + \delta_{3,k} \phi^{\alpha(2+k)\dot\alpha(k+1)}
\Psi^\alpha, \\
0 &=& D \pi^{\alpha(2+k)\dot\alpha(2+k)} - e_{\beta\dot\beta}
\pi^{\alpha(2+k)\beta\dot\alpha(2+k)\dot\beta} + a_{4,k}
e_\beta{}^{\dot\alpha} B^{\alpha(2+k)\beta\dot\alpha(k+1)}, \nonumber
\\
 && + a_{4,k} e^\alpha{}_{\dot\beta} 
B^{\alpha(1+k)\dot\alpha(k+2)\dot\beta} + b_{3,k} e^{\alpha\dot\alpha}
\pi^{\alpha(1+k)\dot\alpha(k+1)} \nonumber \\
 && + \gamma_{4,k} \phi^{\alpha(2+k)\beta\dot\alpha(k+2)} \Psi_\beta
+ \delta_{4,k} \phi^{\alpha(2+k)\dot\alpha(k+1)} \Psi^{\dot\alpha}
\nonumber \\
 && + \gamma_{4,k} \phi^{\alpha(2+k)\dot\alpha(k+2)\dot\beta}
\Psi_{\dot\beta} + \delta_{4,k} \phi^{\alpha(1+k)\dot\alpha(k+2)}
\Psi^\alpha. 
\end{eqnarray}
There are no any possible field redefinitions here, so the requirement
that this deformation to be consistent produces a unique result:
$$
\gamma_{1,k} = \gamma_1, \qquad
\gamma_{2,k} = \frac{\epsilon}{4}\gamma_1, \qquad
\gamma_{3,k} = \frac{3}{4}\gamma_1, \qquad
\gamma_{4,k} = \frac{\epsilon}{2}\gamma_1
$$
$$
\delta_{1,k} = - \frac{\epsilon M}{(k+4)}\gamma_1, \qquad
\delta_{2,k} = - \frac{(k+5)M}{4(k+1)(k+2)}\gamma_1
$$
$$
\delta_{3,k} = - \frac{3\epsilon(k+5)M}{4(k+3)(k+4)}\gamma_1, \qquad
\delta_{4,k} = - \frac{(k+5)M}{2(k+2)(k+3)}\gamma_1
$$
Then, by consistency, we can unambiguously promote this deformation
onto the sector of the auxiliary fields (which are directly connected
with the gauge invariant zero-forms):
\begin{eqnarray}
0 &=& D \Omega^{\alpha(2)} + \frac{M^2}{2}
e^\alpha{}_{\dot\beta} H^{\alpha\dot\beta} + M E^\alpha{}_\beta
B^{\alpha\beta} + 2M^2 E^{\alpha(2)} \varphi - E_{\beta(2)}
 W^{\alpha(2)\beta(2)} \nonumber \\
 && + \epsilon M\gamma_1 \Phi^\alpha \Psi^\alpha 
- \frac{M}{2}\gamma_1 e^\alpha{}_{\dot\alpha} \phi^\alpha
\Psi^{\dot\alpha} + \gamma_1 e_{\beta\dot\alpha}
Y^{\alpha(2)\beta} \Psi^{\dot\alpha}, \nonumber \\ 
0 &=& DB^{\alpha(2)} + M \Omega^{\alpha(2)} + \frac{M}{2} 
e^\alpha{}_{\dot\beta}\pi^{\alpha\dot\beta} - e_{\beta\dot\alpha}
 B^{\alpha(2)\beta\dot\alpha} \\
 && - \epsilon M\gamma_1 \phi^\alpha \Psi^\alpha +
\frac{\epsilon}{4}\gamma_1 Y^{\alpha(2)\beta} \Psi_\beta +
\frac{3}{4}\gamma_1
\phi^{\alpha(2)\dot\alpha} \Psi_{\dot\alpha}, \nonumber \\
0 &=& D\pi^{\alpha\dot\alpha} + M^2 H^{\alpha\dot\alpha} + M
(e_\beta{}^{\dot\alpha} B^{\alpha\beta} + e^\alpha{}_{\dot\beta}
B^{\dot\alpha\dot\beta}) + M^2 e^{\alpha\dot\alpha} \varphi 
- e_{\beta\dot\beta} \pi^{\alpha\beta\dot\alpha\dot\beta} \nonumber \\
 && - M\gamma_1 (\phi^\alpha \Psi^{\dot\alpha} + \phi^{\dot\alpha}
\Psi^\alpha) + \frac{\epsilon}{2}\gamma_1
(\phi^{\alpha\beta\dot\alpha} \Psi_\beta
+ \phi^{\alpha\dot\alpha\dot\beta} \Psi_{\dot\beta}), \nonumber
\end{eqnarray}
and further onto the sector of the physical fields:
\begin{eqnarray}
0 &=& DH^{\alpha\dot\alpha} + e_\beta{}^{\dot\alpha}
\Omega^{\alpha\beta} + e^\alpha{}_{\dot\beta}
\Omega^{\dot\alpha\dot\beta} + M e^{\alpha\dot\alpha} A \nonumber \\
 && + \gamma_1 (\Phi^\alpha \Psi^{\dot\alpha} + \Phi^{\dot\alpha}
\Psi^\alpha), \nonumber \\
0 &=& DA + 2(E_{\alpha(2)} B^{\alpha(2)} + E_{\dot\alpha(2)}
B^{\dot\alpha(2)}) + M e_{\alpha\dot\alpha} H^{\alpha\dot\alpha} \\
 && + \frac{\epsilon}{2}\gamma_1 (\Phi^\alpha \Psi_\alpha +
\Phi^{\dot\alpha} \Psi_{\dot\alpha}) - \frac{3}{2}\gamma_1
e_{\alpha\dot\alpha} (\phi^\alpha \Psi^{\dot\alpha} +
\phi^{\dot\alpha} \Psi^\alpha), \nonumber \\
0 &=& D\varphi + e_{\alpha\dot\alpha} \pi^{\alpha\dot\alpha} + M A
\nonumber \\
 && - \frac{\epsilon}{2}\gamma_1 (\phi^\alpha \Psi_\alpha +
\phi^{\dot\alpha} \Psi_{\dot\alpha}). \nonumber
\end{eqnarray}

\subsection{Deformation for massive spin 3/2}

Here the most general ansatz for the unfolded equations of the gauge
invariant zero-forms looks as follows:
\begin{eqnarray}
0 &=& D Y^{\alpha(3+k)\dot\alpha(k)} - e_{\beta\dot\beta}
Y^{\alpha(3+k)\beta\dot\alpha(k)\dot\beta} + c_{1,k}
e^\alpha{}_{\dot\beta} \phi^{\alpha(2+k)\dot\alpha(k)\dot\beta}
+ d_{1,k} e^{\alpha\dot\alpha} Y^{\alpha(2+k)\dot\alpha(k-1)}
\nonumber \\
 && + \alpha_{1,k} W^{\alpha(3+k)\beta\dot\alpha(k)} \Psi_\beta
 + \beta_{1,k} W^{\alpha(3+k)\dot\alpha(k-1)} \Psi^{\dot\alpha}
\nonumber \\
 && + \alpha_{2,k} B^{\alpha(3+k)\dot\alpha(k)\dot\beta}
\Psi_{\dot\beta} + \beta_{2,k} B^{\alpha(2+k)\dot\alpha(k)}
\Psi^\alpha,  \\
0 &=& D \phi^{\alpha(2+k)\dot\alpha(k+1)} - e_{\beta\dot\beta}
\phi^{\alpha(2+k)\beta\dot\alpha(k+1)\dot\beta} + c_{2,k}
e_\beta{}^{\dot\alpha} Y^{\alpha(2+k)\beta\dot\alpha(k)} \nonumber \\
 && + c_{3,k} e^\alpha{}_{\dot\beta} 
\phi^{\alpha(1+k)\dot\alpha(k+1)\dot\beta} + d_{2,k}
e^{\alpha\dot\alpha} \phi^{\alpha(1+k)\dot\alpha(k)} \nonumber \\
 && + \alpha_{3,k} B^{\alpha(2+k)\beta\dot\alpha(k+1)} \Psi_\beta
+ \beta_{3,k} B^{\alpha(2+k)\dot\alpha(k)} \Psi^{\dot\alpha} \nonumber
\\
 && + \alpha_{4,k} \pi^{\alpha(2+k)\dot\alpha(k+1)\dot\beta}
\Psi_{\dot\beta} + \beta_{4,k} \pi^{\alpha(1+k)\dot\alpha(k+1)}
\Psi^\alpha. 
\end{eqnarray}
Here the solution also appears to be unique:
$$
\alpha_{1,k} = \alpha_1, \qquad
\alpha_{2,k} = \epsilon\alpha_1, \qquad
\alpha_{3,k} = \alpha_1, \qquad
\alpha_{4,k} = \epsilon\alpha_1,
$$
$$
\beta_{1,k} = \frac{\epsilon M}{(k+1)}\alpha_1, \qquad
\beta_{2,k} = \frac{kM}{(k+3)(k+4)}\alpha_1 
$$
$$
\beta_{3,k} = \frac{\epsilon kM}{(k+1)(k+2)}\alpha_1, \qquad
\beta_{4,k} = \frac{kM}{(k+2)(k+3)}\alpha_1 
$$
Promoting these results onto the sector of the physical fields we
obtain:
\begin{eqnarray}
0 &=& D \Phi^\alpha + \epsilon M e^\alpha{}_{\dot\alpha}
\Phi^{\dot\alpha} + 2M E^\alpha{}_\beta \phi^\beta - E_{\beta(2)}
Y^{\alpha\beta(2)} \nonumber \\
 && - \alpha_1 \Omega^{\alpha\beta} \Psi_\beta + \epsilon\alpha_1
e_{\beta\dot\beta} B^{\alpha\beta} \Psi^{\dot\beta} - \epsilon
M\alpha_1 H^{\alpha\dot\alpha} \Psi_{\dot\alpha} - \frac{M}{2}\alpha_1
A \Psi^\alpha + \frac{3\epsilon M}{2}\alpha_1 e^\alpha{}_{\dot\alpha}
 \varphi \Psi^{\dot\alpha}, \nonumber \\
0 &=& D \phi^\alpha + M \Phi^\alpha + \epsilon M 
e^\alpha{}_{\dot\alpha} \phi^{\dot\alpha} - e_{\beta\dot\alpha}
\phi^{\alpha\beta\dot\alpha}  \\
 && + \epsilon \alpha_1 \pi^{\alpha\dot\alpha} \Psi_{\dot\alpha} 
+ \alpha_1 B^{\alpha\beta} \Psi_\beta + \frac{M}{2}\alpha_1 \varphi
\Psi^\alpha. \nonumber 
\end{eqnarray}

\section{Superblock $(3/2, 1)$}

Here we apply the same procedure to the second superblock $(3/2,1)$

\subsection{Deformation for massive spin 3/2}

The most general ansatz for the unfolded equations of the gauge
invariant zero-forms looks like:
\begin{eqnarray}
0 &=& D Y^{\alpha(3+k)\dot\alpha(k)} - e_{\beta\dot\beta}
Y^{\alpha(3+k)\beta\dot\alpha(k)\dot\beta} + c_{1,k}
e^\alpha{}_{\dot\beta} \phi^{\alpha(2+k)\dot\alpha(k)\dot\beta}
+ d_{1,k} e^{\alpha\dot\alpha} Y^{\alpha(2+k)\dot\alpha(k-1)}
\nonumber \\
 && + \alpha_{1,k} \tilde{B}^{\alpha(3+k)\dot\alpha(k)\dot\beta}
\Psi_{\dot\beta} + \beta_{1,k} \tilde{B}^{\alpha(2+k)\dot\alpha(k)}
\Psi^\alpha, \nonumber  \\
0 &=& D \phi^{\alpha(2+k)\dot\alpha(k+1)} - e_{\beta\dot\beta}
\phi^{\alpha(2+k)\beta\dot\alpha(k+1)\dot\beta} + c_{2,k}
e_\beta{}^{\dot\alpha} Y^{\alpha(2+k)\beta\dot\alpha(k)} \nonumber \\
 && + c_{3,k} e^\alpha{}_{\dot\beta} 
\phi^{\alpha(1+k)\dot\alpha(k+1)\dot\beta} + d_{2,k}
e^{\alpha\dot\alpha} \phi^{\alpha(1+k)\dot\alpha(k)} \\
 && + \alpha_{2,k} \tilde{B}^{\alpha(2+k)\beta\dot\alpha(k+1)}
\Psi_\beta + \beta_{2,k} \tilde{B}^{\alpha(2+k)\dot\alpha(k)}
\Psi^{\dot\alpha} \nonumber  \\
 && + \alpha_{3,k} \tilde\pi^{\alpha(2+k)\dot\alpha(k+1)\dot\beta}
\Psi_{\dot\beta} + \beta_{3,k} \tilde\pi^{\alpha(1+k)\dot\alpha(k+1)}
\Psi^\alpha. \nonumber 
\end{eqnarray}
Here also the solution turns out to be unique:
$$
\alpha_{1,k} = \alpha_1, \qquad
\alpha_{2,k} = - \frac{\epsilon}{3}\alpha_1 , \qquad
\alpha_{3,k} = \frac{2}{3}\alpha_1
$$
$$
\beta_{1,k} = - \frac{\epsilon M}{(k+3)}\alpha_1, \qquad
\beta_{2,k} = - \frac{(k+4)M}{3(k+1)(k+2)}\alpha_1, \qquad
\beta_{3,k} = \frac{2(k+4)M}{9(k+2)(k+3)}\alpha_1
$$
Then for the physical fields we obtain:
\begin{eqnarray}
0 &=& D \Phi^\alpha + \epsilon M e^\alpha{}_{\dot\alpha}
\Phi^{\dot\alpha} + 2M E^\alpha{}_\beta \phi^\beta - E_{\beta(2)}
Y^{\alpha\beta(2)} \nonumber \\
 && + \alpha_1 e_{\beta\dot\beta} \tilde{B}^{\alpha\beta}
\Psi^{\dot\beta} + \frac{\epsilon M}{2}\alpha_1 \tilde{A} \Psi^\alpha
+ M\alpha_1 e^\alpha{}_{\dot\alpha} \tilde\varphi \Psi^{\dot\alpha},
\nonumber \\
0 &=& D \phi^\alpha + M \Phi^\alpha + \epsilon M 
e^\alpha{}_{\dot\alpha} \phi^{\dot\alpha} - e_{\beta\dot\alpha}
\phi^{\alpha\beta\dot\alpha} \\
 && + \frac{2}{3}\alpha_1 \tilde\pi^{\alpha\dot\alpha}
\Psi_{\dot\alpha} - \frac{\epsilon}{3}\alpha_1 \tilde{B}^{\alpha\beta}
\Psi_\beta - \epsilon M\alpha_1 \tilde\varphi \Psi^\alpha. \nonumber 
\end{eqnarray}

\subsection{Deformation for massive spin 1}

Similarly, we consider the following ansatz for the gauge invariant
zero-forms:
\begin{eqnarray}
0 &=& D \tilde{B}^{\alpha(2+k)\dot\alpha(k)} - e_{\beta\dot\beta}
\tilde{B}^{\alpha(2+k)\beta\dot\alpha(k)\dot\beta} + a_{1,k}
e^\alpha{}_{\dot\beta} \tilde\pi^{\alpha(1+k)\dot\alpha(k)\dot\beta} +
b_{1,k} e^{\alpha\dot\alpha} \tilde{B}^{\alpha(1+k)\dot\alpha(k-1)}
\nonumber \\
 && + \gamma_{1,k} Y^{\alpha(2+k)\beta\dot\alpha(k)} \Psi_\beta +
\delta_{1,k} Y^{\alpha(2+k)\dot\alpha(k-1)} \Psi^{\dot\alpha}
\nonumber \\
 && + \gamma_{2,k} \phi^{\alpha(2+k)\dot\alpha(k)\dot\beta}
\Psi_{\dot\beta} + \delta_{2,k} \phi^{\alpha(1+k)\dot\alpha(k)}
\Psi^\alpha, \nonumber \\
0 &=& D \tilde\pi^{\alpha(1+k)\dot\alpha(1+k)} - e_{\beta\dot\beta}
\tilde\pi^{\alpha(1+k)\beta\dot\alpha(1+k)\dot\beta} + a_{2,k}
e_\beta{}^{\dot\alpha} \tilde{B}^{\alpha(1+k)\beta\dot\alpha(k)} \\
 && + a_{2,k} e^\alpha{}_{\dot\beta} 
\tilde{B}^{\alpha(1+k)\dot\alpha(1+k)\dot\beta} + b_{2,k}
e^{\alpha\dot\alpha} \tilde\pi^{\alpha(k)\dot\alpha(k)} \nonumber \\
 && + \gamma_{3,k} \phi^{\alpha(1+k)\beta\dot\alpha(k+1)} \Psi_\beta +
\delta_{3,k} \phi^{\alpha(1+k)\dot\alpha(k)} \Psi^{\dot\alpha}
\nonumber \\
 && - \gamma_{3,k} \phi^{\alpha(1+k)\dot\alpha(k+1)\dot\beta}
\Psi_{\dot\beta} - \delta_{3,k} \phi^{\alpha(k)\dot\alpha(k+1)}
\Psi^\alpha. \nonumber 
\end{eqnarray}
Once again we obtain a unique solution:
$$
\gamma_{1,k} = \gamma_1, \qquad
\gamma_{2,k} = - \epsilon\gamma_1, \qquad 
\gamma_{3,k} = \gamma_1,
$$
$$
\delta_{1,k} = - \frac{\epsilon M}{(k+1)}\gamma_1, \qquad
\delta_{2,k} = \frac{kM}{(k+2)(k+3)}\gamma_1, \qquad
\delta_{3,k} = - \frac{k\epsilon M}{(k+1)(k+2)}\gamma_1,
$$
which can be promoted onto the sector of the auxiliary fields:
\begin{eqnarray}
0  &=& D \tilde{B}^{\alpha(2)} + \frac{M}{3} e^\alpha{}_{\dot\alpha}
\tilde\pi^{\alpha\dot\alpha} - e_{\beta\dot\beta} 
\tilde{B}^{\alpha(2)\beta\dot\beta} \nonumber \\
 && + \gamma_1 Y^{\alpha(2)\beta} \Psi_\beta - \epsilon\gamma_1 
\phi^{\alpha(2)\dot\alpha} \Psi_{\dot\alpha}, \\
0 &=& D \tilde\pi^{\alpha\dot\alpha} + \frac{M}{2} 
(e_\beta{}^{\dot\alpha} \tilde{B}^{\alpha\beta} + 
e^\alpha{}_{\dot\beta} \tilde{B}^{\dot\alpha\dot\beta})
- e_{\beta\dot\beta} \tilde\pi^{\alpha\beta\dot\alpha\dot\beta}
\nonumber \\
 && + \gamma_1 ( \phi^{\alpha\beta\dot\alpha} \Psi_\beta +
\phi^{\alpha\dot\alpha\dot\beta} \Psi_{\dot\beta}), \nonumber 
\end{eqnarray}
and onto the sector of the physical fields:
\begin{eqnarray}
0 &=& D \tilde{A} + 2(E_{\alpha(2)} \tilde{B}^{\alpha(2)} +
E_{\dot\alpha(2)} \tilde{B}^{\dot\alpha(2)}) \nonumber \\
 && + 2\gamma_1 (\Phi^\alpha \Psi_\alpha - \Phi^{\dot\alpha} 
\Psi_{\dot\alpha}) + 2\epsilon\gamma_1 e_{\alpha\dot\alpha}
(\phi^\alpha \Psi^{\dot\alpha} - \phi^{\dot\alpha} \Psi^\alpha),
\nonumber  \\
0 &=& D \tilde\varphi + e_{\alpha\dot\alpha}
\tilde\pi^{\alpha\dot\alpha} + \frac{M}{2} \tilde{A} \\
 && - \gamma_1 (\phi^\alpha \Psi_\alpha - \phi^{\dot\alpha}
\Psi_{\dot\alpha}). \nonumber
\end{eqnarray}

\end{document}